\documentclass[english]{article}
\usepackage[T1]{fontenc}
\usepackage[utf8]{inputenc}
\usepackage[a4paper]{geometry}
\usepackage{float}
\usepackage{amsmath}
\usepackage{amssymb}
\usepackage{graphicx}
\usepackage{cite}
\usepackage[unicode=true]{hyperref}
\usepackage{setspace}
\DeclareMathOperator{\arccosh}{arccosh}

\begin{document}

\title{Mean First Passage Time of the Symmetric Noisy Voter Model with Arbitrary
Initial and Boundary Conditions}
\author{Rytis Kazakevi\v{c}ius, Aleksejus Kononovicius}

\date{Institute of Theoretical Physics and Astronomy, Vilnius University}

\maketitle
\begin{abstract}
Models of imitation and herding behavior often underestimate
the role of individualistic actions and assume symmetric boundary
conditions. However, real-world systems (e.g., electoral processes)
frequently involve asymmetric boundaries. In this study, we explore how
arbitrarily placed boundary conditions influence the mean first passage time
in the symmetric noisy voter model, and how individualistic behavior
amplifies this asymmetry. We derive exact analytical expressions for mean
first passage time that accommodate any initial condition and two types of
boundary configurations: (i) both boundaries absorbing, and (ii) one
absorbing and one reflective. In both scenarios, mean first passage time
exhibits a clear asymmetry with respect to the initial condition, shaped by
the boundary placement and the rate of independent transitions. Symmetry in
mean first passage time emerges only when absorbing boundaries are
equidistant from the midpoint. Additionally, we show that Kramers' law holds
in both configurations when the rate of independent transitions is large.
Our analytical results are in excellent agreement with numerical
simulations, reinforcing the robustness of our findings.
\end{abstract}

\section{Introduction}

The mean first passage time (abbr.~MFPT) is a tool widely used in
various fields such as quantum statistical mechanics~\cite{Qiu2012,Wu2013,Friedman2017,Liu2020,Hasegawa2022,Kulkarni2023,Kewming2024},
laser physics~\cite{Roy1980,Cao2001}, chemistry~\cite{Kalantar2019,Preston2021,Ravichandir2025},
biology~\cite{Polizzi2016,Zhang2016,Singh2018}. It is particularly
prominent in the research on stochastic processes and random walks
\cite{Zhou2013,Kim2020}, and most recently in the context of target
search problems~\cite{Grebenkov2024}. The MFPT, also known as the
first hitting time or first detection time, refers to the average
time it takes for a system initially in one specific state (or states)
to reach a different specific state (or states) for the first time.
The MFPT can be directly calculated from the first passage time distribution
(abbr.~FPTD)~\cite{Risken1989,Gardiner2004,Grebenkov2024}; however,
for many stochastic processes, the FPTD is known only in the form
of an infinite series~\cite{Borodin2012}. In certain cases, e.g.,
for birth--death processes~\cite{Kononovicius2019BDJStat}, MFPT can
be a used to obtain a reasonable approximation for the FPTD. In sociophysics,
which applies physics-based methods to social phenomena, MFPT is increasingly
used to study aspects of human behavior, such as polarization and
segregation in voting dynamics~\cite{Bassolas2021}, and how the trading
strategies affect the financial markets~\cite{Kutner2019PhysA}. Voter
models, in particular, are employed to simulate social networks and
the influence of mass media on opinion formation~\cite{Hu2024}, statistical
voting patterns~\cite{FernandezGarcia2014PRL,Braha2017PlosOne,Kononovicius2017Complexity,Marmani2020Entropy},
polarization, and consensus formation~\cite{Baronchelli2018RSOS,Bhat2020}.
Comparing voter model predictions against empirical social network
data analysis is also a promising research direction~\cite{Fushimi2010,Kimura2011,Kimura2012,Rawal2019,He2021}.
While there are many other opinion dynamics
models~\cite{Castellano2009RevModPhys,Jedrzejewski2019CRP,Noorazar2020EPJP},
only voter models have been as well-suited for the MFPT
analysis~\cite{Redner2019CRP}.
In this context, the investigation of MFPT within the framework of
voter models constitutes a compelling direction for future research,
with the theoretical and applied potential.

An extensive body of work exists that either analytically derives,
or explores numerically, the MFPT for a range of modified and extended
versions of the voter model~\cite{Masuda2010,Rozanova2017,Redner2019CRP,Bhat2020,Perachia2022}.
Most of these works focus on evaluating the consensus time, which
refers to the average time for a polarized system to reach a consensus
state for the first time~\cite{Masuda2010,Rozanova2017,Redner2019CRP,Perachia2022}.
It has also been applied to quantify other characteristic times, such
as the opinion switching time~\cite{Kudtarkar2024} and opinion polarization
time~\cite{Bhat2020}. Therefore, the analysis of MFPT in voter models
helps researchers understand how fast a population can change, how
this the pace of change depends on system characteristics~\cite{Bhat2020}
and variety of opinion formation mechanisms~\cite{Redner2019CRP}.
By providing a quantitative framework for the expected timing of transitions
between the states, MFPT enables the exploration of communication
patterns, influence, and collective decision-making in networks~\cite{Lee2012,Fushimi2010,Hu2024}.
This exploration could have potential applications for financial risk
management~\cite{Kutner2019PhysA}, as the noisy voter model (often
under the name of Kirman model~\cite{Kirman1993QJE}) has been shown
to be a good model for opinion dynamics in the financial markets as
well~\cite{Alfarano2008Dyncon,Kononovicius2012,Kononovicius2019OB}.

Despite the growing interest in first passage time statistics for
various modifications of the voter models, most studies have mainly
relied on numerical simulations~\cite{Masuda2010,Rozanova2017,Perachia2022,Chen2022}.
Furthermore, many of the earlier works have set out to explore how
the MFPT scales with the system size~\cite{MartinezGarcia2012,Rozanova2017,Redner2019CRP}
but have neglected scaling based on other parameters, such as independent
transition rates, introduced in other common modifications of the
voter model. Notably~\cite{Bhat2020,Kudtarkar2024} have recently
explored MFPT dependence on the independent transition rates. Even
in these works analytical approximations for MFPT have been obtained
only in limited cases, such as for low independent transition rates
\cite{Kudtarkar2024} or for fixed boundary conditions~\cite{Bhat2020}.
Therefore, a general framework that analyzes MFPT dependence on transition
rates and boundary conditions can still be developed further.

To simplify the analytical derivation, the MFPT dependence on the
initial condition can be made symmetric by choosing symmetric boundary
conditions~\cite{Redner2001}. Yet this assumption is highly restrictive,
as real-life problems may have inherent asymmetry. For example, in
many multi-party proportional representation systems, a political
party must surpass a minimum vote threshold, often 4\% or 5\%,
to gain representation in the parliament~\cite{Gallagher2005}. On
the other hand, a party would be considered successful (and likely
to win a multi-party election) if it received more than 30\% of
the votes. Thus, a natural asymmetry in the political expectations
arises. Consequently, in this work, we focus on asymmetry in the MFPT,
as it better captures the non-uniform nature of threshold-based systems
and decision-making processes in social systems. The primary aim of
this study is to explore how asymmetry in boundary conditions influence
asymmetry observed in MFPT and how asymmetric MFPT scales as independent
transition rates grow larger. To facilitate the present analysis,
we instead assume symmetry in the independent transition rates, an
assumption we anticipate relaxing in the future.

This paper is organized as follows. In Section~\ref{sec:nvm}, we
describe our definition and approach to the numerical simulation of
the noisy voter model. In Section~\ref{sec:MFPT}, we present the
main analytical results and compare them against the numerical simulations.
In Section~\ref{sec:Discussion}, we discuss we obtained
results with a focus on their validation and possible applications.
Conclusions and future outlook follow in Section~\ref{sec:Conclusions}.  In
the main body of the text, to keep it more approachable and concise, we have
skipped some of the derivation steps, these are discussed in detail in the
\ref{sec:A} (derivation of the ordinary differential equation for MFPT),
\ref{sec:B} (solution of the ordinary differential equation for MFPT) and
\ref{sec:C} (analysis of a few selected special cases).

\section{Noisy voter model\label{sec:nvm}}

The original formulation of the voter model~\cite{Clifford1973} involved
spatial competition between two species. In the original model, during
every simulation step, a random site on rectangular grid was selected,
and then its contents (a single individual) were replaced by a copy
of an individual from a neighboring site. This copying mechanism was
well received by the opinion dynamics community
\cite{Castellano2009RevModPhys,Liggett1999},
as with a slight change in terminology, it can be seen to be equivalent
to conformist behavior in social response theory~\cite{Willis1965,Nail2016APPA}.
Here, we will consider a generalization of the voter model, which
incorporates not only peer copying behavior but also spontaneous transitions
without interaction with a peer. Because spontaneous transitions are
similar to noise, this generalization is often referred to as the
noisy voter model. This generalization of the voter model was originally
introduced in~\cite{Granovsky1995}, but a remarkably similar model
was independently proposed in~\cite{Kirman1993QJE} with economic
modeling in mind. Thus, some works refer to this model as Kirman's
model or the herding model~\cite{Alfarano2008Dyncon,Ruseckas2011,Kononovicius2012},
though the microscopic foundation of both models is mostly the same.
The noisy voter model is one of the most prevalent models in sociophysics
\cite{Castellano2009RevModPhys}. It exhibits rich phenomenology and
has successfully been used to replicate electoral data across different
countries~\cite{FernandezGarcia2014PRL,Braha2017PlosOne,Kononovicius2017Complexity,Marmani2020Entropy}.

Let us formulate the noisy voter model as follows. Let there be a
fixed number $N$ of particles (also agents or voters) in the simulated
system. Any particle can occupy any of the two available states. Let
us label the states as ``$0$'' and ``$1$''. Transitions between
the states occur either independently (at rate $r_{i}$, where the
index $i$ would match the destination state label) or due to the
interaction with other particles in the destination state (let the
interaction rate between any two particles be $h$). If $X$ denotes
the number of particles in state ``$1$'', then a single transition
will change the $X$ by at most one step (up or down). Under these
assumptions and with the selected notation, the transition rates with
respect to $X$ are given by:
\begin{equation}
\lambda\left(X\rightarrow X+1\right)= \lambda^{+} =
\left(N-X\right)\left(r_{1}+hX\right), \quad
\lambda\left(X\rightarrow X-1\right)= \lambda^{-} =X\left(r_{0}+h\left[N-X\right]\right).
\end{equation}

Because the transition rates encode a one-step process, we employ the
birth--death process formalism~\cite{VanKampen2007NorthHolland} to derive a
stochastic differential equation (abbr. SDE) that approximates the discrete noisy voter model:
\begin{equation}
dx = \frac{\lambda^{+}-\lambda^{-}}{N}dt +
\sqrt{\frac{\lambda^{+}+\lambda^{-}}{N^{2}}}dW \approx
   h\left[\varepsilon_{1}(1-x)-\varepsilon_{0}x\right]dt +
   \sqrt{2hx(1-x) + \frac{h \varepsilon_{1}}{N}(1-x) + \frac{h \varepsilon_{0}}{N}x}dW.
\end{equation}

Here, $N$ denotes the number of agents. The equation above is suitable for
analyzing finite-size effects and related phenomena. However, in this study,
we focus on the large-$N$ regime. Taking the limit $N \rightarrow \infty$
and assuming that the transition rates are much smaller than $N$, the SDE
governing the fraction of agents in state ``$1$'', i.e.,~$x = \tfrac{X}{N}$,
simplifies to
\begin{equation}
dx = h\left[\varepsilon_{1}(1-x) - \varepsilon_{0}x\right]dt + \sqrt{2hx(1-x)}dW. \label{eq:sdex}
\end{equation}

We interpret the SDE above in the It\^o sense, where $W$ denotes the
standard Wiener process. In this formulation, we introduce a dimensionless
notation for the independent transition rates, $\varepsilon_{i} =
\tfrac{r_{i}}{h}$. This choice decouples the overall timescale of the
process (controlled by $h$) from the shape of its dynamics (determined by
$\varepsilon_{i}$).  Thus, without loss of generality, let us set $h=1$ and
consider only the impact of independent transition rates $\varepsilon_{i}$.
For $h\neq 1$, the obtained mean first-passage times would need to be
trivially multiplied by $h$.

Obtaining the steady-state distribution of the process described by
Eq.~\eqref{eq:sdex} is trivial~\cite{Gardiner2004}. It can easily
be shown to be the Beta distribution with respective shape parameters
equal to $\varepsilon_{i}$. The probability density function
of the steady-state distribution is given by:
\begin{equation}
P_{\infty}\left(x\right) = \frac{\Gamma\left(\varepsilon_{0}+\varepsilon_{1}\right)}{\Gamma\left(\varepsilon_{0}\right)\Gamma\left(\varepsilon_{1}\right)}x^{\varepsilon_{1}-1}\left(1-x\right)^{\varepsilon_{0}-1}.
\end{equation}
Notably, steady-state distribution can also be derived for the discrete
noisy voter model with finite $N$. It follows a Beta-binomial distribution
with the number of trials equal to $N$ and the shape parameters given
by $\varepsilon_{i}$ (see Appendix~B in~\cite{Mori2019PRE}). The
independent transition rates, $\varepsilon_{0}$ and $\varepsilon_{1}$,
literally control the shape of the Beta distribution. If both are
less than $1$, $P_{\infty}\left(x\right)$ is bimodal (i.e.,~the
distribution has a U-shape). If both are greater than $1$, $P_{\infty}\left(x\right)$
is unimodal (i.e.,~the distribution has a bell-shape). If $\varepsilon_{0}=\varepsilon_{1}=1$,
then the Beta distribution reduces to the uniform distribution. If
the values of the independent transition rates differ, $P_{\infty}\left(x\right)$
will become skewed to the right (if $\varepsilon_{1}>\varepsilon_{0}$)
or to the left (if $\varepsilon_{1}<\varepsilon_{0}$). In this work,
we focus on a symmetric noisy voter model, i.e.,~$\varepsilon_{0}=\varepsilon_{1}=\varepsilon$.

We numerically solve Eq.~\eqref{eq:sdex}, with $h=1$ and $\varepsilon_{0}=\varepsilon_{1}=\varepsilon$,
using Euler--Maruyama method~\cite{Kloeden1992Springer}. Starting
from the initial condition $x_{0}$, we iterate the following difference
equation (in it $\Delta t$ is the constant time step, and $\xi_{i}$
are standard normal random variables),
\begin{equation}
x_{i+1}=x_{i}+\varepsilon\left(1-2x_{i}\right)\Delta
t+\sqrt{2x_{i}\left(1-x_{i}\right)\Delta t}\xi_{i}, \label{eq:num-sol-diff}
\end{equation}
until either lower boundary $L$ or higher boundary $H$ is passed
(we require that $L\leq x_{0}\leq H$). If that boundary is absorbing,
we record the time, $T= \left(i+1\right)\Delta t$, as a new sample
of the first passage time and restart the simulation. If the boundary
condition is reflective, then the new $x_{i+1}$ value is clipped
to the range of valid values (e.g.,~if after iterating the difference
equation we would have $x_{i+1}\leq L$, we instead set $x_{i+1}=L$).
To constrain the simulation, we require that any first passage time
would be at most $T_{\text{max}}$. If $\left(i+1\right)\Delta t>T_{\text{max}}$
without reaching an absorbing boundary condition, we ignore the current
run and restart the simulation. Unless noted otherwise, the numerical
results reported here have been obtained with $\Delta t=10^{-5}$,
$T_{\text{max}}=10^{6}$ and by taking $10^{5}$ samples.
Alternatively, a variable time step method~\cite{Ruseckas2016} or
higher-order methods~\cite{Kloeden1992Springer} (e.g.,~the Milstein
approximation method) could be used; however, these methods would require
more computational resources without producing a qualitative impact on the
obtained results.

The code implementing the described approach to sampling the MFPT
from the symmetric noisy voter model, along with the relevant scripts
reproducing all the figures presented in the later sections, is available
on GitHub~\cite{github}.

\section{The mean first passage time\label{sec:MFPT}}

\subsection{General solution}

If first-passage time distribution (abbr.~FPTD), $p_{T}\left(T\right)$, of the stochastic process has a closed-form
expression, mean first-passage time (abbr.~MFPT) can be calculated directly
\begin{equation}
\overline{T}=\int_{0}^{\infty}Tp_{T}\left(T\right)dT.\label{eq:MFPT-definition}
\end{equation}
However, for many stochastic processes, the FPTD can be expressed
analytically only as an infinite sum, rather than as a closed-form
expression~\cite{Borodin2012}. Even when the FPTD can be expressed
in terms of elementary or special functions, the integral above often
remains analytically intractable and must be evaluated numerically
\cite{Risken1989}. Therefore, a more commonly employed approach in
physics to obtain the MFPT is to use the time-backward Fokker--Planck
equation.

For the symmetric noisy voter model described by Eq.~\eqref{eq:sdex},
with $h=1$ and $\varepsilon_{0}=\varepsilon_{1}=\varepsilon$, the
corresponding Fokker--Planck is
\begin{equation}
\frac{\partial}{\partial t}P\left(x,t\big|x_{0},0\right)=-\varepsilon\frac{\partial}{\partial x}\left[\left(1-2x\right)P\left(x,t\big|x_{0},0\right)\right]+\frac{\partial^{2}}{\partial x^{2}}\left[x\left(1-x\right)P\left(x,t\big|x_{0},0\right)\right].\label{eq:fokker-planck-nvm-x}
\end{equation}
It can be shown that the time-backward Fokker--Planck equation~\cite{Gardiner2004}
associated with the Fokker--Planck equation above is
\begin{equation}
\frac{\partial}{\partial t}P\left(x,t\big|x_{0},0\right)=\varepsilon\left(1-2x_{0}\right)\frac{\partial}{\partial x_{0}}P\left(x,t\big|x_{0},0\right)+x_{0}\left(1-x_{0}\right)\frac{\partial^{2}}{\partial x_{0}^{2}}P\left(x,t\big|x_{0},0\right).\label{eq:back-fokker-planck-nvm-x}
\end{equation}
In the~\ref{sec:A},
Eqs.~\eqref{eq:Time-back-ward-F-P-E-AG}--\eqref{eq:MFPT-ODE-gen},
we have shown that the time-backward Fokker--Planck equation above
leads to the following ordinary differential equation (abbr.~ODE)
for the MFPT:
\begin{equation}
x_{0}\left(1-x_{0}\right)\frac{d^{2}}{dx_{0}^{2}}\overline{T}+\varepsilon\left(1-2x_{0}\right)\frac{d}{dx_{0}}\overline{T}=-1.\label{eq:ODE-gen-eps-main}
\end{equation}
To derive the ODE above from the Fokker--Planck equation, we assumed that
the parameter $\varepsilon$ is time-independent. This is a common assumption
in the
literature~\cite{Granovsky1995,Castellano2009RevModPhys,Noorazar2020EPJP}.
Although, some recent works have shown that for certain problems it is
convenient to treat $\varepsilon$ as being
time-dependent~\cite{Kononovicius2024PhysA}. Consequently,
ODE~\eqref{eq:ODE-gen-eps-main} is not applicable to scenarios involving
time-dependent transition rates.
For a general framework to derive ODEs for the MFPT see Refs.~\cite{Risken1989,Gardiner2004}. The obtained equation for the MFPT is a second-order ODE and must
be supplemented by two boundary conditions. Therefore, the MFPT problem
is well-poised only as an escape from a bounded domain problem~\cite{Redner2001}.
This approach cannot be applied to problems with only one absorbing
boundary condition. In such cases, alternative approaches can be taken,
such as perturbation techniques~\cite{Grasman1999,Lindner2004,Moreira2015,Urdapilleta2015},
Laplace transforms~\cite{Ahmad2019}, or a direct solution of backward
Fokker--Planck equation~\cite{Kim2020}. In this paper, we obtain a
general solution of the Eq.~\eqref{eq:ODE-gen-eps-main} (see~\ref{sec:B}
for more details), which specifies the MFPT for the symmetric noisy
voter model.

Let the symmetric noisy voter model process start at $x_{0}$. Let
the boundary conditions be placed at $L$ and $H$, so that $0\leq L\leq x_{0}\leq H\leq1$.
These boundary conditions confine the process to a clipped interval
$\left[L,H\right]$ instead of the natural interval $\left[0,1\right]$.
Then in case of two absorbing boundary conditions, we obtain the MFPT (see~\ref{sec:B} for derivation)
\begin{equation}
\overline{T}_{LH}=\frac{\overline{T}_{p}\left(H\right)\beta_{1-L}-\overline{T}_{p}\left(L\right)\beta_{1-H}}{\beta_{1-H}-\beta_{1-L}}+\frac{\overline{T}_{p}\left(L\right)-\overline{T}_{p}\left(H\right)}{\beta_{1-H}-\beta_{1-L}}\beta_{1-x_{0}}+\overline{T}_{p}\left(x_{0}\right).\label{eq:mfpt-lh-aa}
\end{equation}
In the above $\beta_{z}=\beta_{z}\left(1-\varepsilon,1-\varepsilon\right)$
is the incomplete beta function (the frequently repeated arguments
have been omitted for brevity), and $\overline{T}_{p}\left(z\right)$
is a particular solution of Eq.~\eqref{eq:ODE-gen-eps-main}, which
is given by
\begin{equation}
\overline{T}_{p}\left(z\right)=\frac{1}{\Gamma\left(2\varepsilon\right)\Gamma\left(1-\varepsilon\right)}G_{3,3}^{2,3}\left(z\Bigg|\begin{array}{c}
1,1,2\left(1-\varepsilon\right)\\
1,1-\varepsilon,0
\end{array}\right).
\end{equation}
Here $\Gamma\left(\varepsilon\right)$ is the gamma function, and
$G_{p,q}^{m,n}\left(z\Bigg|\begin{array}{c}
a_{1},a_{2},..,a_{p}\\
b_{1},b_{2},..,b_{q}
\end{array}\right)$ is Meijer~$G$-function.

If the boundary condition at $L$ is reflective instead (this choice,
in the context of the voter model, could be interpreted as representing
an inflexible voter base~\cite{Mobilia2007JStatMech,Galam2007PhysA,Kononovicius2014PhysA,Khalil2018PRE}),
while the boundary condition at $H$ remains absorbing, the MFPT is
given by (see~\ref{sec:B} for derivation)
\begin{equation}
\overline{T}_{r}= \left[\frac{\Gamma\left(\varepsilon\right)\Gamma\left(\varepsilon\right)}{\Gamma\left(2\varepsilon\right)}-\beta_{L}\left(\varepsilon,\varepsilon\right)\right]\left[\beta_{1-x_{0}}\left(1-\varepsilon,1-\varepsilon\right)-\beta_{1-H}\left(1-\varepsilon,1-\varepsilon\right)\right]+\overline{T}_{p}\left(x_{0}\right)-\overline{T}_{p}\left(H\right).\label{eq:mfpt-lh-ra}
\end{equation}
If boundary condition at $H$ would be reflective (this choice, in
the context of the voter model, could be interpreted as representing
an inflexible opposition~\cite{Mobilia2007JStatMech,Galam2007PhysA,Kononovicius2014PhysA,Khalil2018PRE}),
and boundary condition at $L$ would be absorbing, the above solution
would still apply with just swapping of the $L$ and $H$ symbols.

\subsection{\texorpdfstring{Obtaining reduced mean first passage time expressions for certain $\varepsilon$}{Obtaining reduced mean first passage time expressions for certain epsilon}}

Eqs.~\eqref{eq:mfpt-lh-aa} and \eqref{eq:mfpt-lh-ra} constitute
the main analytical results of this paper. At first glance, these
MFPT expressions may appear problematic to evaluate due to the presence
of special functions. However, for certain values of the independent
transition rate, these special functions may be reduced to expressions
involving basic functions. In particular, significant simplification
occurs when $\varepsilon=\tfrac{n}{2}$ with $n\in\mathbb{N}$, due
to the mathematical properties of the special functions involved.
Notably, special attention should be given to the cases with integer
$\varepsilon$, as in these cases the standard relation between the
incomplete beta function and the hyper-geometric function breaks down.
For the integer $\varepsilon$, $\beta_{1-z}\left(1-\varepsilon,1-\varepsilon\right)$,
$\beta_{z}\left(\varepsilon,\varepsilon\right)$ and $\overline{T}_{p}\left(z\right)$
must be evaluated directly from Eqs.~\eqref{eq:T_HL_eps-N} and \eqref{eq:Tvid_LR_HA-1}.
Thus, let us consider the first few values, i.e., $\varepsilon=0$,
$\tfrac{1}{2}$, $1$, $\tfrac{3}{2}$, $2$ and $\tfrac{5}{2}$.

\begin{figure}[t]
\centering{}\includegraphics[width=1\textwidth]{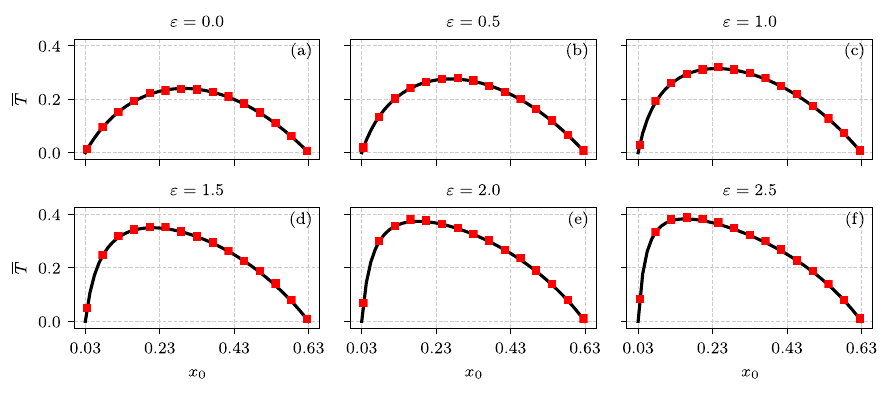}\caption{The mean first passage time $\overline{T}$ dependence on the initial
condition $x_{0}$ with various values of independent transition rate
$\varepsilon$ for the case with absorbing boundary conditions at
$L$ and $H$. The black curves correspond to Eq.~\eqref{eq:mfpt-lh-aa},
while the red squares represent estimates obtained by numerical simulation
of Eq.~\eqref{eq:sdex} with $h=1$ and $\varepsilon_{0}=\varepsilon_{1}=\varepsilon$.
Boundary conditions were placed at $L=0.03$ and $H=0.63$.\protect\label{fig:mfpt-aa}}
\end{figure}

\begin{figure}[t]
\centering{}\includegraphics[width=1\textwidth]{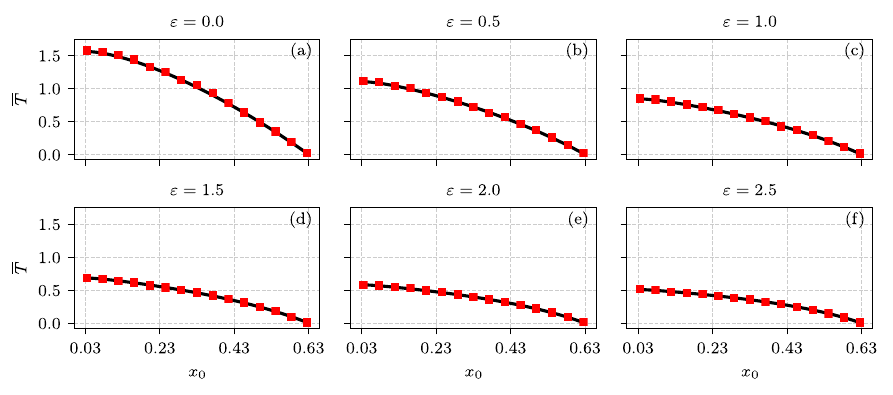}\caption{The mean first passage time $\overline{T}$ dependence on the initial
condition $x_{0}$ with various values of independent transition rate
$\varepsilon$ for the case with reflective boundary condition at
$L$ and absorbing boundary condition at $H$. The black curves correspond
to Eq.~\eqref{eq:mfpt-lh-ra}, while the red squares represent estimates
obtained by numerical simulation of Eq.~\eqref{eq:sdex} with $h=1$
and $\varepsilon_{0}=\varepsilon_{1}=\varepsilon$. In all instances
boundary conditions were placed at $L=0.03$ and $H=0.63$.\protect\label{fig:mfpt-ra}}
\end{figure}

For the considered $\varepsilon$ values, one of the parametrizations
of the incomplete beta function can be reduced to
\begin{equation}
\beta_{1-z}\left(1-\varepsilon,1-\varepsilon\right)=\begin{cases}
1-z, & \text{for }\varepsilon=0,\\
2\arcsin\left(\sqrt{1-z}\right), & \text{for }\varepsilon=\frac{1}{2},\\
\ln\left(1-z\right) & \text{for }\varepsilon=1,\\
\frac{2\left(1-2z\right)}{\sqrt{z\left(1-z\right)}}, & \text{for }\varepsilon=\frac{3}{2},\\
\frac{1-2z}{z\left(1-z\right)}+2\left(\ln\left(z\right)-\ln\left(1-z\right)\right), & \text{for }\varepsilon=2,\\
\frac{2\left(16z^{3}-24z^{2}+6z+1\right)}{3z^{3/2}\left(1-z\right)^{3/2}}, & \text{for }\varepsilon=\frac{5}{2}.
\end{cases}\label{eq:beta-reduced}
\end{equation}
Similarly, the other parametrization can be reduced to
\begin{equation}
\beta_{z}\left(\varepsilon,\varepsilon\right)=\begin{cases}
\ln\left(z\right), & \text{for }\varepsilon=0,\\
2\arcsin\left(\sqrt{z}\right), & \text{for }\varepsilon=\frac{1}{2},\\
z, & \text{for }\varepsilon=1,\\
\frac{1}{4}\left(\left(2z-1\right)\sqrt{z\left(1-z\right)}+\arcsin\left(\sqrt{z}\right)\right), & \text{for }\varepsilon=\frac{3}{2},\\
\frac{1}{6}\left(3-2z\right)z^{2}, & \text{for }\varepsilon=2,\\
\frac{1}{64}\left(\left(16z^{3}-24z^{2}+2z+3\right)\sqrt{z\left(1-z\right)}+3\arcsin\left(\sqrt{z}\right)\right). & \text{for }\varepsilon=\frac{5}{2}.
\end{cases}\label{eq:beta-reduced-2}
\end{equation}
 Likewise, the particular solution can be reduced to
\begin{equation}
\overline{T}_{p}\left(z\right)=\begin{cases}
\left(z-1\right)\ln\left(1-z\right)-z\ln\left(z\right), & \text{for }\varepsilon=0,\\
\frac{1}{2}\left(\pi^{2}+\left(2\arccosh\left(\sqrt{z}\right)\right)^{2}\right), & \text{for }\varepsilon=\frac{1}{2},\\
\ln\left(z\right)-\ln\left(1-z\right) & \text{for }\varepsilon=1,\\
\frac{1}{2}\left(\frac{\left(2z-1\right)\arccos\left(\sqrt{z}\right)}{\sqrt{z\left(1-z\right)}}-1\right), & \text{for }\varepsilon=\frac{3}{2},\\
\frac{1}{6}\left(\frac{z}{z-1}+2\ln\left(1-z\right)\right), & \text{for }\varepsilon=2,\\
\frac{\left(20z^{2}-20z-3\right)\sqrt{z\left(1-z\right)}-3\left(16z^{3}-24z^{2}+6z+1\right)\arccos\left(\sqrt{z}\right)}{96z^{3/2}\left(1-z\right)^{3/2}}, & \text{for }\varepsilon=\frac{5}{2}.
\end{cases}\label{eq:particular-reduced}
\end{equation}
Relevant derivations for a few of these special cases are provided
in ~\ref{sec:C}.
The reduced expressions enable a more transparent analytical treatment
of the MFPT dependence on the initial condition. These expressions
also facilitate a comparison between analytical expressions and results
of the numerical simulation (see Figs.~\ref{fig:mfpt-aa} and \ref{fig:mfpt-ra}).

Figures~\ref{fig:mfpt-aa} and \ref{fig:mfpt-ra} show an excellent
agreement between the obtained MFPT expressions and numerical simulation
results for the two distinct boundary problems. For the both considered
boundary problems, the dependence of MFPT on the independent transition
rate is highly asymmetric. For the case with two absorbing boundaries
(see Fig.~\ref{fig:mfpt-aa}), the degree of asymmetry increases
with $\varepsilon$ as the MFPT peak shifts away from the middle of
the allowed $x$ value range towards the boundary condition which
lies closer to the natural boundary. Also, the maximum MFPT grows
larger with $\varepsilon$. For the case with one reflective and one
absorbing boundary (see Fig.~\ref{fig:mfpt-ra}), the maximum MFPT
always lies at the reflective boundary, but the maximum $\overline{T}$
decreases with $\varepsilon$. Though, as will be shown in Section~\ref{subsec:kramers-law},
the dependence of $\overline{T}$ on $\varepsilon$ in both considered
cases is not as trivial as can be observed from Figs.~\ref{fig:mfpt-aa}
and \ref{fig:mfpt-ra}.

\subsection{\texorpdfstring{Detailed analysis of the $\varepsilon=0$ case}
{Detailed analysis of the epsilon=0 case}}
Here, let us consider $\varepsilon=0$ case in more detail. This case
yields the most compact and tractable expressions for the MFPT within
our framework. This case also proves particularly interesting as it
coincides with the original voter model on a complete graph (i.e.,
if any particle may interact with any other particle).

Let us insert appropriate replacements for the incomplete beta function,
taken from Eqs.~\eqref{eq:beta-reduced} and \eqref{eq:beta-reduced-2},
and the particular solution, Eq.~\eqref{eq:particular-reduced},
into Eq.~\eqref{eq:mfpt-lh-aa}. Then Eq.~\eqref{eq:mfpt-lh-aa}
simplifies to
\begin{equation}
\overline{T}_{LH}\left(x_{0}\big|\varepsilon=0\right)=\frac{Lf\left(H\right)-Hf\left(L\right)}{H-L}+\frac{f\left(L\right)-f\left(H\right)}{H-L}x_{0}+f\left(x_{0}\right).\label{eq:mfpt-lh-aa-eps-zero}
\end{equation}
Here we have introduced a placeholder function
\begin{equation}
f\left(z\right)=\left(z-1\right)\ln\left(1-z\right)-z\ln\left(z\right),
\end{equation}
which allows us to greatly simplify the MFPT expression above. Notably,
it can be shown that MFPT dependence on the initial condition reaches
maximum value at
\begin{equation}
x_{0,\text{max}}=\max\left[\overline{T}_{LH}\left(x_{0}\big|\varepsilon=0\right)\right]=\frac{1}{2}\left(1-\tanh\left[\frac{f\left(H\right)-f\left(L\right)}{2\left(H-L\right)}\right]\right).
\end{equation}

The meaning of the placeholder function might not be immediately obvious,
but if we let the absorbing boundaries coincide with the natural boundaries,
i.e., $L=0$ and $H=1$, MFPT becomes
\begin{equation}
\overline{T}_{01}\left(x_{0}\big|\varepsilon=0\right)=\left(x_{0}-1\right)\ln\left(1-x_{0}\right)-x_{0}\ln\left(x_{0}\right)=f\left(x_{0}\right).
\end{equation}
In the literature, MFPT to the natural boundaries of the voter model
is commonly referred to as the mean consensus time~\cite{Redner2001}.
It represents the average time required for all particles to end up
in the same state (either \textquotedblleft $0$\textquotedblright{}
or \textquotedblleft $1$\textquotedblright ), thereby achieving consensus.
Notably, the expression for the mean consensus time is identical to
$f\left(x_{0}\right)$; hence, the placeholder function above represents
this quantity. The analytical expression for the mean consensus time
has been derived in numerous previous studies and is well known to
involve logarithmic terms across various extensions of the voter model
\cite{Redner2001,Sood2005,Sood2008,Baxter2011,Mukhopadhyay2020}.
Although, to the best of our knowledge, the generalized MFPT expression,
Eq.~\eqref{eq:mfpt-lh-aa-eps-zero}, has not yet been studied in
detail. Therefore, we continue with the analysis of the generalized
MFPT expression, which reveals how the symmetry or asymmetry of the
MFPT can be controlled through the choice of boundary conditions.

The meaning of the first term in Eq.~\eqref{eq:mfpt-lh-aa-eps-zero}
becomes obvious by inserting $x_{0}=L$ or $x_{0}=H$ into the equation.
Namely, the first term ensures that MFPT will be zero at the boundary
conditions $L$ and $H$.

The meaning of the second term in Eq.~\eqref{eq:mfpt-lh-aa-eps-zero}
is revealed by choosing symmetric boundary conditions, i.e., $L=\tfrac{1}{2}-d$
and $H=\tfrac{1}{2}+d$ (with $0<d\leq\tfrac{1}{2}$). With these
boundary conditions, the second term disappears, which allows us to
conclude that this term is responsible for the asymmetry in the MFPT.
The term disappears because the placeholder (consensus time) function
is symmetric in respect to the midpoint, i.e., $f\left(\tfrac{1}{2}-z\right)=f\left(\tfrac{1}{2}+z\right)$
(with $0\leq z\leq\tfrac{1}{2}$) or, alternatively, $f\left(z\right)=f\left(1-z\right)$
(with $0\leq z\leq1$). Thus, if the boundary conditions are placed
symmetrically to the midpoint, MFPT will also be symmetric in respect
to the midpoint,
\begin{equation}
\overline{T}_{LH}\left(x_{0}\big|\varepsilon=0\right)=\overline{T}_{LH}\left(1-x_{0}\big|\varepsilon=0\right),\qquad\text{if }L=\frac{1}{2}-d\text{ and }H=\frac{1}{2}+d.
\end{equation}
The asymmetry in the MFPT arises from the
nonlinear, position-dependent diffusion coefficient in the Fokker--Planck
equation, Eq.~\eqref{eq:fokker-planck-nvm-x}. Specifically, the
diffusion term $x\left(1-x\right)$ increases from zero at $x=0$,
reaches its maximum at $x=\tfrac{1}{2}$, and then goes back to zero
at $x=1$. This means that the rate of diffusion is not uniform across
the interval. If one boundary is placed near a region of slower diffusion
(close to $0$ or $1$) and the other near a region of greater diffusion
(closer to $\tfrac{1}{2}$), the particle is more likely to reach
the latter boundary first. For instance, if $L$ is close to $0$
and $H=\tfrac{1}{2}$, and the initial condition $x_{0}$ lies between
them, the particle tends to reach $H$ faster because the diffusion
coefficient is significantly larger close to $H$ than near $L$.
This contrasts with the standard Wiener process, where the diffusion
coefficient is constant. In that case, when the process is bounded
in the interval $\left[L,H\right]$, the MFPT is symmetric with respect
to the midpoint $\tfrac{H+L}{2}$, regardless of the absolute placement
of the boundaries.

However, in real-life applications, especially for social processes,
symmetric boundary problems are quite rare. For example, in many proportional
representation systems, a political party must surpass a minimum vote
threshold, often $4\%$ or $5\%$ , to gain representation in parliament
\cite{Gallagher2005}. Though, to win an election a party does not
need to gain $95\%$ or $96\%$ of the votes, as usually top parties
in democratic multi-party elections gain roughly $30\%$ to $40\%$
of votes. Thus, some important real-life problems involves asymmetric
boundaries. Moreover, sociopolitical dynamics, such as incumbency
advantage~\cite{Erikson2015}, media influence and strategic voting
\cite{Cohen2009CommRes}, further skew the outcome distribution. Thus,
we analyze a more general asymmetric problem instead.

For the reflective lower boundary, given $L>0$, we have 
\begin{equation}
\overline{T}_{r}\left(x_{0}\big|\varepsilon=0\right)=\left(H-x_{0}\right)g\left(L\right)-f\left(H\right)+f\left(x_{0}\right),
\end{equation}
where $g\left(z\right)$ is a placeholder function given by
\begin{equation}
g\left(z\right)=\ln\left(1-z\right)-\ln\left(z\right)=2\arctan\left(1-2z\right).
\end{equation}
It effectively encodes the attraction towards the reflective boundary
(i.e., it accounts for the time wasted moving towards the reflective
boundary). Due to this attraction, the maximum MFPT, for the case
with reflective boundary at $L$, is located at the reflective boundary
$L$.

Note that the discussion surrounding the case with reflective boundary
at $L$ above applies only when $L>0$. Similarly, if the reflective
boundary condition would be placed at $H$ instead, then the discussion
would apply only for $H<1$. This is because a reflective boundary
at $L=0$ (or $H=1$) would not work as intended. Given that $\varepsilon=0$,
it is obvious that the drift term disappears from the Fokker--Planck
equation, Eq.~\eqref{eq:fokker-planck-nvm-x}, while the diffusion
term goes to zero at the natural boundaries, $x=0$ and $x=1$. Consequently,
at the natural boundaries there would be no force (deterministic or
stochastic) to push the process away from it, and the process would
get stuck (or, in other words, absorbed) at the natural boundary instead
of the being reflected.

\subsection{\texorpdfstring{Analysis of the $\varepsilon=\tfrac{1}{2}$ case}{Analysis of the epsilon=1/2 case}}
This case also yields compact and tractable MFPT expressions within
our approach. Let us insert appropriate replacements for the incomplete
beta function, taken from Eqs.~\eqref{eq:beta-reduced} and \eqref{eq:beta-reduced-2},
and the particular solution, Eq.~\eqref{eq:particular-reduced},
into Eq.~\eqref{eq:mfpt-lh-aa}. Then for two absorbing boundaries,
we have that
\begin{equation}
\overline{T}_{LH}\left(x_{0}\big|\varepsilon=\frac{1}{2}\right)=\frac{1}{2}\left[\arcsin\left(2H-1\right)-\arcsin\left(2x_{0}-1\right)\right]\cdot\left[\arcsin\left(2x_{0}-1\right)-\arcsin\left(2L-1\right)\right].
\end{equation}
It can be shown that the MFPT maximum, in the case with two absorbing
boundaries, is located at
\begin{equation}
x_{0,\text{max}}=\max\left[\overline{T}_{LH}\left(x_{0}\big|\varepsilon=\frac{1}{2}\right)\right]=\frac{1}{2}\left[1+\sin\left(\frac{\arcsin\left(2H-1\right)+\arcsin\left(2L-1\right)}{2}\right)\right].
\end{equation}
From the above it is trivial to see that with symmetric absorbing
boundaries, $L=\tfrac{1}{2}-d$ and $H=\tfrac{1}{2}+d$ (with $0<d\leq\tfrac{1}{2}$),
the MFPT maximum would be located at the midpoint.

By inserting the appropriate replacements from Eqs.~\eqref{eq:beta-reduced}
-- \eqref{eq:particular-reduced} into Eq.~\eqref{eq:mfpt-lh-ra},
we obtain the MFPT dependence with reflective lower boundary, $L>0$,
\begin{align}
\overline{T}_{r}\left(x_{0}\big|\varepsilon=\frac{1}{2}\right) & =\frac{1}{2}\left[\arcsin\left(2H-1\right)-\arcsin\left(2x_{0}-1\right)\right]\nonumber \\
 & \quad\cdot\left[\arcsin\left(2H-1\right)-2\arcsin\left(2L-1\right)+\arcsin\left(2x_{0}-1\right)\right].
\end{align}
Once again the MFPT maximum when one reflective boundary would be
located at the reflective boundary.

\begin{figure}[t]
\begin{centering}
\includegraphics[width=1\textwidth]{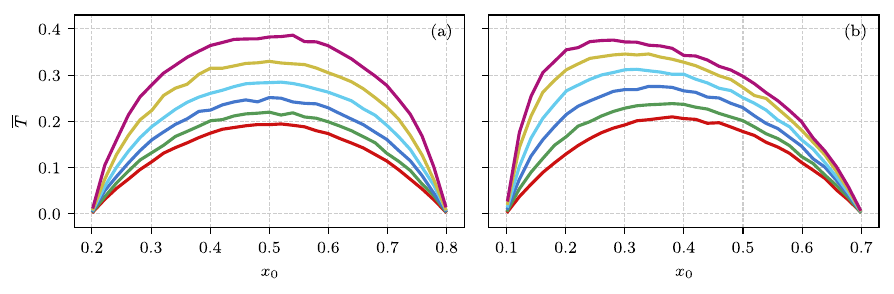}
\par\end{centering}
\caption{The mean first passage time dependence $\overline{T}$ on the initial
condition $x_{0}$ with symmetric absorbing boundaries (a), and with
asymmetric absorbing boundaries (b). Colored curves show the MFPT
dependence obtained by numerical simulation with different rates:
$\varepsilon=0$ (red curves), $0.8$ (green), $1.6$ (blue), $2.4$
(cyan), $3.2$ (yellow), $4.0$ (magenta). Boundary conditions for
the symmetric case (subfigure (a)) were placed at $L=0.2$ and $H=0.8$,
for the asymmetric case (subfigure (b)) they were placed at $L=0.1$
and $H=0.7$.\protect\label{fig:sym-asym}}
\end{figure}

Using similar reasoning as in the previous subsection, one can show
that 
\begin{equation}
\overline{T}_{LH}\left(x_{0}\big|\varepsilon=\frac{1}{2}\right)=\overline{T}_{LH}\left(1-x_{0}\big|\varepsilon=\frac{1}{2}\right),\qquad\text{if }L=\frac{1}{2}-d\text{ and }H=\frac{1}{2}+d.
\end{equation}
The condition necessary to obtain MFPT symmetry is the same as in
the $\varepsilon=0$ case. As shown in~\ref{sec:C},
when $\varepsilon=1$ or $\varepsilon=\tfrac{3}{2}$, choosing symmetric
boundaries also yields symmetric MFPT. In principle, the same approach
can be applied to any $\varepsilon=\tfrac{n}{2}$, with $n\in\mathbb{N}$,
since the MFPT expressions involve trigonometric, logarithmic, and
polynomial functions, all of which exhibit inherent symmetries. We
therefore speculate that this symmetry might hold for arbitrary values
of $\varepsilon$. While we haven\textquoteright t been able to prove
this analytically, analysis by numerical simulation supports our intuition
(see Fig.~\ref{fig:sym-asym}~(a)).

\begin{figure}[t]
\begin{centering}
\includegraphics[width=0.5\textwidth]{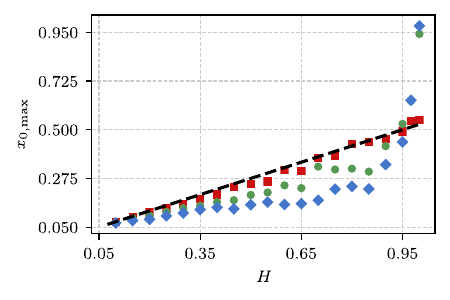}
\par\end{centering}
\caption{Numerically simulated dependence of $x_{0,\text{max}}$ on the placement
of the asymmetric absorbing boundary conditions. Lower boundary is
fixed at $L=0.05$, while the higher boundary $H$ is treated as an
independent variable. Dashed black line highlights the midpoint between
the boundary conditions, $\tfrac{H+L}{2}$.
Colored symbols show numerical
simulation results obtained with different rates: $\varepsilon=0$
(red squares), $1.6$ (green circles), $3.2$ (blue diamonds).\label{fig:x0max}}
\end{figure}

Conducting the same numerical analysis with asymmetric boundary conditions
(see Fig.~\ref{fig:sym-asym}~(b)) seems to indicate that $x_{0,\text{max}}$
tends towards boundary condition, which is closest to the natural
boundary condition. In Fig.~\ref{fig:sym-asym}~(b), the lower boundary
condition is closer to the natural boundary condition ($0.1$ is closer
to $0$, than $0.7$ to $1$), thus we observe that the MFPT maximum
shifts to the lower values as $\varepsilon$ increases. This behavior
can be examined in more detail by fixing $L$ and exploring how $x_{0,\text{max}}$
depends on the location of the higher boundary condition, $H$. Results
of the additional numerical analysis, shown in Fig.~\ref{fig:x0max},
quantitatively confirm the qualitative intuition obtained from inspecting
Fig.~\ref{fig:sym-asym}~(b). In other words, $x_{0,\text{max}}$
indeed tends towards the boundary condition closer to the natural
boundary condition (all colored curves are below the dashed midpoint
line for $H<0.95$, and are above the dashed midpoint line for $H>0.95$),
while the deviation from the midpoint grows larger as $\varepsilon$
increases.

\subsection{\texorpdfstring{Mean first passage time scaling dependence on $\varepsilon$\label{subsec:kramers-law}}{Mean first passage time scaling dependence on}}
MFPT scaling behavior can also be reduced to a simpler expression
by considering the dependence on $\varepsilon$ in the large $\varepsilon$
limit. In~\ref{sec:A},
Eqs.~\eqref{eq:Tvid_LR_HA} -- \eqref{eq:T-r-large-eps-approx},
we have be shown that Eq.~\eqref{eq:mfpt-lh-ra} can be approximated
by
\begin{equation}
\overline{T}\left(\varepsilon\right)\simeq\frac{2\sqrt{\pi}}{2^{2\varepsilon}\varepsilon^{3/2}}\cdot\frac{H}{\left(1-H\right)^{\varepsilon}H^{\varepsilon}}=\frac{2\sqrt{\pi}H}{\varepsilon^{3/2}}\mathrm{e}^{\varepsilon\ln\left[\frac{1}{4H\left(1-H\right)}\right]}.\label{eq:mfpt-lh-ra-large-eps-approx}
\end{equation}
In~\ref{sec:A}, we demonstrate that Eq.~\eqref{eq:mfpt-lh-ra} can be
approximated by the expression above. This result is obtained by performing
a series expansion of Euler beta and incomplete beta functions, specifically
in the large-$\varepsilon$ limit. The expansion assumes that the influence
of the lower boundary is negligible. In the above, the symbol $\simeq$
implies functional dependence
of $\overline{T}$ on $\varepsilon$. As this is functional dependence,
it is not normalized in respect to $L$ and $x_{0}$ (only terms involving
$\varepsilon$ are present).

Large $\varepsilon$ limit implies that the drift term in the Fokker--Planck
equation, Eq.~\ref{eq:fokker-planck-nvm-x}, dominates the diffusion
term. Stiff potential rendering the effects of a random force negligible.
Consequently, in this limit, the nonlinearity of the diffusion coefficient
can be safely ignored in the MFPT analysis. The functional dependence
above, in its purer form, i.e., 
\begin{equation}
\overline{T}\sim\varepsilon^{-3/2}\mathrm{e}^{\varepsilon},\label{eq:kramers-law}
\end{equation}
matches Kramers' law for the MFPT in a stiff harmonic potential~\cite{Kim2015}.

\begin{figure}[t]
\centering{}\includegraphics[width=1\textwidth]{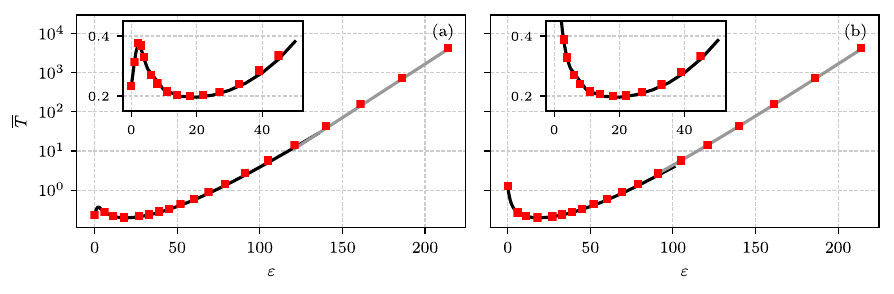}\caption{The mean first passage time $\overline{T}$ dependence on the independent
transition rate $\varepsilon$: (a) the case with absorbing boundary
conditions at $L$ and $H$, (b) the case with reflective boundary
condition at $L$ and absorbing boundary condition at $H$. The black
curves correspond to Eq.~\eqref{eq:mfpt-lh-aa} (for subfigure~(a))
and Eq.~\eqref{eq:mfpt-lh-ra} (for subfigure~(b)), the gray curves
follow Kramers' law, Eq.~\eqref{eq:kramers-law}. The red squares
represent estimates obtained by numerical simulation of Eq.~\eqref{eq:sdex}
with $h=1$ and $\varepsilon_{0}=\varepsilon_{1}=\varepsilon$. The
insets show a more detailed dependence in the range of smaller $\varepsilon$.
The boundary and initial conditions were set as follows $L=0.03$,
$H=0.63$ and $x_{0}=0.23$.\protect\label{fig:epsi-mfpt}}
\end{figure}

A comparison between the numerical simulations and analytical predictions
in Fig.~\ref{fig:epsi-mfpt} confirms that the Kramers' law (gray
curves) holds for both cases: with two absorbing boundary conditions
(subfigure (a)) and with one reflective and one absorbing boundary
condition (subfigure (b)). For small values of $\varepsilon$, the
dependence of $\overline{T}$ on $\varepsilon$ is accurately described
by Eq.~\eqref{eq:mfpt-lh-aa} (black curve in subfigure (a)) and
Eq.~\eqref{eq:mfpt-lh-ra} (black curve in subfigure (b)). The insets
in Fig.~\ref{fig:epsi-mfpt} provide a closer view of the dependence
in the small $\varepsilon$ value range.

\section{Discussion\label{sec:Discussion}}
\subsection{Numerical validation of analytical results and relation to other noisy voter models}
We have derived exact analytical expressions for the mean first passage
time (abbr.~MFPT), $\overline{T}$, in the symmetric noisy voter model (i.e.,
with identical independent transition rates, $\varepsilon_{0}=\varepsilon_{1}=\varepsilon$).
Eq.~\eqref{eq:mfpt-lh-aa} corresponds to the case where both boundary
conditions (located at $L$ and $H$) are absorbing, while Eq.~\eqref{eq:mfpt-lh-aa}
applies when one of the boundaries is reflective (without loss of
generality, we assume $L$ is reflective). Unlike previous works,
our expressions accommodate arbitrary placements of the initial (denoted
by $x_{0}$) and boundary conditions, provided they satisfy a reasonable
constraint $0\leq L\leq x_{0}\leq H\leq1$. We have verified that
the analytical results are in excellent quantitative agreement with
numerical simulations.

In models involving imitation or herding behavior, the role of individualistic
behavior is frequently underestimated~\cite{Kudtarkar2024}. Nevertheless,
our research reveals that arbitrarily placed boundary conditions make
$\overline{T}$ dependence on $x_{0}$ asymmetric, and the presence
of individualistic behavior can significantly exaggerate this asymmetry.
When both boundaries are absorbing, asymmetry increases with $\varepsilon$
(see Figs.~\ref{fig:mfpt-aa}, \ref{fig:sym-asym}~(b) and \ref{fig:x0max}):
the location of $\overline{T}$ maximum shifts toward the absorbing
boundary that lies closer to the natural boundary, and the value of
the maximum MFPT rises. In contrast, when one boundary is reflective,
the qualitative asymmetry remains largely unchanged (see Fig.~\ref{fig:mfpt-ra}),
while the overall MFPT decreases with increasing $\varepsilon$. Though,
$\overline{T}$ dependence on $\varepsilon$ for fixed $x_{0}$ suggests
that the behavior across a broader range of $\varepsilon$ is more
complex than it appears from the analysis of moderate $\varepsilon$
(see Fig.~\ref{fig:epsi-mfpt}). For extremely large $\varepsilon$,
MFPT scales according to the Kramers' law for stiff harmonic potential
\cite{Kim2015}. This contrasts with the asymptotic MFPT behavior
found for the preferential voter model, which exhibits exponential
decay~\cite{Lee2012}.

We have also conducted a detailed analysis of the case with $\varepsilon=0$,
i.e., when the noisy voter model reduces to the original voter model
on complete graph. We showed that the MFPT is symmetric only when
the boundaries are placed symmetrically around the point where the
diffusion is highest---at $x=\tfrac{1}{2}$ (see Fig.~\ref{fig:sym-asym}~(a)).
This symmetry breaks down when the boundaries are not evenly placed
around this point, leading to a directional bias in the MFPT due to
the spatial variation in diffusivity (see Fig.~\ref{fig:sym-asym}~(b)).
There are numerous previous works, which have considered mean consensus
or mean polarization times for the original voter model and its minor
modifications~\cite{Redner2001,Sood2005,Sood2008,Baxter2011,Mukhopadhyay2020}.
Mean consensus times often correspond to the average transition time
from a polarized state (e.g., $x=\tfrac{1}{2}$) to a complete consensus
state (i.e., $x=0$ or $x=1$). Mean polarization times solve the
opposite problem, i.e., finding the average time from consensus state
to the polarized state. We show that these are just special cases
of our more general MFPT expression, Eq.~\eqref{eq:mfpt-lh-aa-eps-zero}.
Previous studies~\cite{Redner2001,Sood2005,Sood2008,Baxter2011,Mukhopadhyay2020}
have established that these MFPTs exhibit a logarithmic dependence
on the initial condition. Consistently, our expression reduces to
a logarithmic dependence in the limiting case of $\varepsilon=0$.
However, in the general case, for $\varepsilon>0$, general expressions
involve special functions, which often are not as trivial to evaluate.
We show that, for $\varepsilon=\tfrac{n}{2}$ with $n\in\mathbb{N}$,
the MFPT can still be expressed in terms of trigonometric, logarithmic
functions and polynomials. Previous work by Bhat and Redner~\cite{Bhat2020}
established this result for values up to $n=4$. Our analysis extends
these findings by revealing that such expressions arise from more
general special functions, which we have carefully incorporated into
our study.

\subsection{Future Real-World Applications in Social and Physical Systems}
In real-world social and political systems, boundary conditions are rarely
symmetric. For example, proportional representation systems impose a lower
vote threshold of about 4--5\% for parliamentary entry~\cite{Gallagher2005},
while a party is already considered successful with roughly 30--40\% of the
    vote (far from the 95\% upper bound assumed in symmetric models). Additional sociopolitical forces such as incumbency advantage~\cite{Erikson2015}, media influence, and strategic voting~\cite{Cohen2009CommRes} further introduce asymmetry. Consequently, we focus on asymmetric boundary conditions when analyzing MFPT, as this better reflects threshold-based decision processes in social systems.

A key theoretical contribution of this work is the exact analytical solution
of the ordinary differential equation~\eqref{eq:ODE-gen-eps-main}, expressed
in terms of the incomplete beta function and the Meijer~$G$-function, as
presented in ~\ref{sec:B} (Eq.~\eqref{eq:MFPT-gen}). To the best of our
knowledge, this solution is both unique and previously undocumented in the
physical or social science literature. Additionally,
Eqs.~\eqref{eq:mfpt-lh-aa} and Eqs.~\eqref{eq:mfpt-lh-ra}, which incorporate
arbitrary boundary placements, represent a novel generalization of earlier
models. These expressions extend prior studies by accommodating
non-symmetric and mixed boundary conditions, offering a more flexible and
realistic framework for analyzing MFPT in noisy voter dynamics. Prior
studies often impose symmetric boundaries to simplify analysis
\cite{Castellano2009RevModPhys,Jedrzejewski2019CRP,Redner2019CRP,Noorazar2020EPJP},
but this assumption is overly restrictive given the inherently uneven
expectations in systems with thresholds. The noisy voter model is well
suited for social applications such as opinion dynamics. For example, a
party's stable core support may be modeled by a reflective boundary at low $x$, while a funding threshold is represented by an absorbing boundary. Thus, our analytical MFPT result is directly actionable for quantifying how long a new political party is expected to take to reach viability under stochastic dynamics.

Beyond social systems, the model also has relevance in physical contexts. It
has been shown that the stochastic differential equation~\eqref{eq:sdex} can
be approximated by heterogeneous diffusion processes (abbr.~HDPs)
\cite{Kazakevicius2021PRE}, which describe particle diffusion in systems
with temperature gradients~\cite{Cherstvy2013NJPhys,Kazakevicius2015JStat}
and external potentials~\cite{Kazakevicius2016}. These connections suggest
potential applications in anomalous diffusion
\cite{Cherstvy2013NJPhys,Kazakevicius2016}, power-law statistics
\cite{Kazakevicius2015JStat,Kazakevicius2015ChaosSF} , and $1/f$ noise
\cite{Kazakevicius2014PhysicaA, Ruseckas2016}. Conversely, HDPs can also be
approximated by voter models, allowing the MFPT expressions derived here to
serve as useful estimates in physical systems. Anomalous diffusion has also
been observed in social dynamics, such as human movement intermittency
\cite{Luo2025PhysicaA} and parliamentary presence
\cite{Kononovicius2017Complexity,Kononovicius2020JStatMech}. Nonlinear
transformations of the noisy voter model can mimic statistical properties of
fractional Brownian motion~\cite{Kononovicius2022CSF,Kazakevicius2021PRE}, a
standard model for persistent and anti-persistent diffusion and long-range
memory~\cite{Ibe2013}. This research may therefore be useful for studying
persistence and anti-persistence in various time-series data. In the case of
time-dependent herding, approximate formulas for the first-passage time
distribution have been obtained under restrictive conditions, such as small
signal intensity and neglecting one boundary~\cite{Mukhopadhyay2020}. Thus
the chosen direction, toward more realistic and flexible modeling of
asymmetry and time dependence, appears especially promising, as it builds directly on the analytical foundations established in this work and opens pathways for broader applications.

Looking ahead, we plan to extend this framework to asymmetric noisy voter models, particularly in regimes with asymmetric transition rates. These models are analytically challenging but offer rich dynamics that merit further exploration. We also recognize the potential for incorporating time-dependent parameters and boundary conditions to model adaptive or externally driven systems. While exact solutions may not be feasible in such cases, approximate methods and numerical simulations could provide valuable insights.

\section{Conclusions\label{sec:Conclusions}}

We derived exact analytical expressions for the mean first passage time
(abbr.~MFPT) in the symmetric noisy voter model with identical transition
rates. These results apply to both absorbing and mixed boundary conditions
and allow for arbitrary placements of the initial and boundary states within
the interval $[0,1]$. Our analytical predictions show excellent agreement
with numerical simulations. Our analysis highlights how individualistic
behavior and asymmetric boundary placement significantly influence MFPT.
These factors introduce directional bias and shift the location of the
maximum MFPT. For large noise levels ($\varepsilon$), the MFPT scales
according to Kramers' law, in contrast to the exponential decay observed in preferential voter models.

In absence of individualistic behaviour ($\varepsilon=0$), the model reduces to the classical voter model, where MFPT exhibits a logarithmic dependence on the initial condition. Symmetry in MFPT is preserved only when boundaries are placed evenly around $x=\frac{1}{2}$, the point of highest diffusivity. Otherwise, asymmetry emerges due to spatial variation in diffusion. Consensus and polarization times, widely studied in earlier works, emerge as special cases of our more general MFPT formulation. For specific values of $\varepsilon=\tfrac{n}{2}$, MFPTs can be expressed using elementary functions, extending previous results to broader parameter ranges. These expressions offer practical tools for analyzing MFPTs in heterogeneous diffusion processes, which can be approximated by voter models and vice versa.

A central novelty of our study is the general solution to ordinary
differential equation~\eqref{eq:ODE-gen-eps-main}, presented in \ref{sec:B}
(Eq.~\eqref{eq:MFPT-gen}), which involves the incomplete beta function and
the Meijer~$G$-function. To the best of our knowledge, this solution is both unique and previously undocumented in physical or social science literature. Additionally, the MFPT results in Eqs.~\eqref{eq:mfpt-lh-aa} and Eqs.~\eqref{eq:mfpt-lh-ra}, which explicitly incorporate boundary conditions, represent a significant generalization of earlier models and introduce new analytical structures to the study of stochastic dynamics.

Finally, we acknowledge that model simplifications (such as symmetric transition rates and fixed boundaries) affect the generality of the results. However, these choices are necessary to obtain tractable analytical solutions. Future work may explore more complex scenarios, including asymmetric time-dependent transition rates, external perturbations, and dynamic boundary conditions.

{\small\begin{singlespace}

\newpage\appendix

\section{Derivation of the ordinary differential equation for the mean first
passage time\label{sec:A}}

Here, we derive an ordinary differential equation for calculating
the mean first-passage time from the Fokker--Planck equation in the
single-variable case. For the multivariable case, a general framework
for deriving MFPT ODEs can be found in Refs.~\cite{Risken1989,Grasman1999,Gardiner2004}.

The survival probability---the probability that a particle remains
within the interval $x\in\left[a,b\right]$ up to time $t$ (smaller
or equal to first passage time), without having reached the interval
boundary---is given by:
\begin{equation}
\mathrm{Prob}\left(T\geq t\right)=G\left(x_{0},t\right)=\int_{a}^{b}P\left(x,t\big|x_{0},0\right)dx,\label{eq:Survival-function}
\end{equation}
where, $P\left(x,t\mid x_{0},0\right)$ is the transition probability
density, i.e., the conditional probability that a particle starting
at position $x_{0}$ at time zero is found at position $x$ at time
$t$. The probability that the particle has exited the domain $x\in\left[a,b\right]$
by time $t$ is given by:
\begin{equation}
\mathrm{Prob}\left(T\leq t\right)=1-\mathrm{Prob}\left(T\geq t\right)=1-G\left(x_{0},t\right).
\end{equation}

The FPTD, denoted $p_{T}\left(T\right)$, describes the probability
that the particle crosses either boundary $a$ or $b$ for the first
time within the infinitesimal interval $\left(t,t+dt\right)$. It
is given by the negative time derivative of the survival probability:
\begin{equation}
p_{T}\left(t\right)=\frac{\partial}{\partial t}\mathrm{Prob}\left(T\leq t\right)=\frac{\partial}{\partial t}\left(1-G\left(x_{0},t\right)\right)=-\frac{\partial}{\partial t}G\left(x_{0},t\right).
\end{equation}

The MFPT is defined as the expected value of the first-passage time:
\begin{equation}
\overline{T}\left(x_{0}\right)=\int_{0}^{\infty}tp_{T}\left(t\right)dt=-\int_{0}^{\infty}t\frac{\partial}{\partial t}G\left(x_{0},t\right)dt.\label{eq:MFPT-definition-A}
\end{equation}
To evaluate the MFPT, we consider the integral:
\begin{equation}
I=-\int_{0}^{\infty}t\frac{\partial}{\partial t}G\left(x_{0},t\right)dt.\label{eq:Int_I}
\end{equation}
Let us apply the integration by parts using the standard formula:
\begin{equation}
\int udv=uv-\int vdu.\label{eq:vu-dvu}
\end{equation}
Here, let us choose
\begin{equation}
u=t\quad\Rightarrow\quad du=dt,\qquad dv=\frac{\partial}{\partial t}G\left(x_{0},t\right)dt\quad\Rightarrow\quad v=G\left(x_{0},t\right).\label{eq:variables}
\end{equation}
Substituting Eqs.~ \eqref{eq:vu-dvu} and \eqref{eq:variables}
into the formula~\eqref{eq:Int_I}, we obtain:
\begin{equation}
I=-\int_{0}^{\infty}tdG\left(x_{0},t\right)=-\left[tG\left(x_{0},t\right)\Big|_{0}^{\infty}-\int_{0}^{\infty}G\left(x_{0},t\right)dt\right].
\end{equation}
By definition, $G\left(x_{0},0\right)=1$ for all $x_{0}\in\left[a,b\right]$.
This immediately implies that $tG\left(x_{0},t\right)\to0$ as $t\to0$
, since the factor of $t$ vanishes. To ensure that the boundary term
in an integration by parts vanishes at the upper limit, we require
$\lim_{t\to\infty}tG\left(x_{0},t\right)=0$. This condition holds
if $G\left(x_{0},t\right)$ decays faster than $\tfrac{1}{t}$ as
$t\to\infty$. For instance, this is satisfied when $G\left(x_{0},t\right)\sim e^{-t}$
or $G\left(x_{0},t\right)\sim\tfrac{1}{t^{\gamma}}$ with $\gamma>1$.
Assuming such decay, the boundary term $tG\left(x_{0},t\right)\Big|_{0}^{\infty}$
vanishes, and we are left with:
\begin{equation}
I=-\int_{0}^{\infty}t\frac{\partial}{\partial t}G\left(x_{0},t\right)dt=\int_{0}^{\infty}G\left(x_{0},t\right)dt.\label{eq:Final-integral}
\end{equation}
Comparing formulas~\eqref{eq:MFPT-definition-A} and~\eqref{eq:Final-integral},
we find the relationship between the MFPT and the survival probability:
\begin{equation}
\overline{T}\left(x_{0}\right)=-\int_{0}^{\infty}t\frac{\partial}{\partial t}G\left(x_{0},t\right)dt=\int_{0}^{\infty}G\left(x_{0},t\right)dt.\label{eq:MFPT-relation-to-G(x0,t)}
\end{equation}

If the system is governed by the Fokker--Planck equation
\begin{equation}
\frac{\partial}{\partial t}P\left(x,t\big|x_{0},0\right)=-\frac{\partial}{\partial x}A\left(x\right)P\left(x,t\big|x_{0},0\right)+\frac{1}{2}\frac{\partial^{2}}{\partial x^{2}}B\left(x\right)P\left(x,t\big|x_{0},0\right),\label{eq:F-P-E-gen}
\end{equation}
and the process is time-homogeneous---that is, $P\left(x,t\big|x_{0},0\right)=P\left(x,0\big|x_{0},-t\right)$,
then the corresponding time-backward Fokker--Planck equation takes
the form
\begin{equation}
\frac{\partial}{\partial t}P\left(x,t\big|x_{0},0\right)=A\left(x_{0}\right)\frac{\partial}{\partial x_{0}}P\left(x,t\big|x_{0},0\right)+\frac{1}{2}B\left(x_{0}\right)\frac{\partial^{2}}{\partial x_{0}^{2}}P\left(x,t\big|x_{0},0\right).\label{eq:Time-back-ward-F-P-E-AG}
\end{equation}
To relate this to the survival probability, we integrate both sides
of the time-backward equation~\eqref{eq:Time-back-ward-F-P-E-AG}
over $x\in\left[a,b\right]$:
\begin{equation}
\frac{\partial}{\partial t}\int_{a}^{^{b}}P\left(x,t\big|x_{0},0\right)dx=A\left(x_{0}\right)\frac{\partial}{\partial x_{0}}\int_{a}^{^{b}}P\left(x,t\big|x_{0},0\right)dx+\frac{1}{2}B\left(x_{0}\right)\frac{\partial^{2}}{\partial x_{0}^{2}}\int_{a}^{^{b}}P\left(x,t\big|x_{0},0\right)dx.\label{eq:Time-backward-F-P-E-integral}
\end{equation}
Recognizing that the integral on the left-hand side corresponds to
the survival probability $G\left(x_{0},t\right)$, as defined in Eq.~\eqref{eq:Survival-function},
we can rewrite the equation as:
\begin{equation}
\frac{\partial}{\partial t}G\left(x_{0},t\right)=A\left(x_{0}\right)\frac{\partial}{\partial x_{0}}G\left(x_{0},t\right)+\frac{1}{2}B\left(x_{0}\right)\frac{\partial^{2}}{\partial x_{0}^{2}}G\left(x_{0},t\right).
\end{equation}
This shows that the survival probability $G\left(x_{0},t\right)$
satisfies the time-backward Fokker--Planck equation and can be obtained
by directly solving it. Next, we integrate both sides of this equation
with respect to $t$ from $0$ to $\infty$
\begin{equation}
\int_{0}^{\infty}\frac{\partial}{\partial t}G\left(x_{0},t\right)dt=A\left(x_{0}\right)\frac{\partial}{\partial x_{0}}\int_{0}^{\infty}G\left(x_{0},t\right)dt+\frac{1}{2}B\left(x_{0}\right)\frac{\partial^{2}}{\partial x_{0}^{2}}\int_{0}^{\infty}G\left(x_{0},t\right)dt,
\end{equation}
Recalling the previously obtained relation between the MFPT and the
survival probability, Eq.~\eqref{eq:Survival-function}, we
see that the equation above is
\begin{equation}
G\left(x_{0},\infty\right)-G\left(x_{0},0\right)=A\left(x_{0}\right)\frac{d}{dx_{0}}\overline{T}\left(x_{0}\right)+\frac{1}{2}B\left(x_{0}\right)\frac{d^{2}}{dx_{0}^{2}}\overline{T}\left(x_{0}\right).\label{eq:MFPT-ODE-0}
\end{equation}
We assume that $G\left(x,\infty\right)=0$, since any particle is
eventually absorbed. From the initial condition $p\left(x,0\big|x_{0},0\right)=\delta\left(x_{0}-x\right)$,
it follows that $G\left(x,0\right)=1$ for $x\in\left[a,b\right]$.
Substituting these values into the Eq.~\eqref{eq:MFPT-ODE-0} yields:
\begin{equation}
\frac{1}{2}B\left(x_{0}\right)\frac{d^{2}}{dx_{0}^{2}}\overline{T}\left(x_{0}\right)+A\left(x_{0}\right)\frac{d}{dx_{0}}\overline{T}\left(x_{0}\right)=-1,\label{eq:MFPT-ODE-gen}
\end{equation}
which is the ODE that arises from the time-backward Fokker--Planck
equation for the MFPT, Eq.~\eqref{eq:Time-back-ward-F-P-E-AG}.

\section{Solution of the ordinary differential equation governing the mean
first-passage time\label{sec:B}}

We now solve the ODE that determines the MFPT. The symmetric noisy
voter model is governed by the time-backward Fokker--Planck equation,
Eq.~\eqref{eq:back-fokker-planck-nvm-x}. Combining this with the
previously established relation given by Eq.~\eqref{eq:MFPT-ODE-gen}
yields the following ODE for the MFPT:
\begin{equation}
\frac{d^{^{2}}}{dx_{0}^{2}}\overline{T}\left(x_{0}\right)+\varepsilon\frac{1-2x_{0}}{x_{0}\left(1-x_{0}\right)}\frac{d}{dx_{0}}\overline{T}\left(x_{0}\right)=-\frac{1}{x_{0}\left(1-x_{0}\right)}.\label{eq:ODE-gen-eps}
\end{equation}
First, we find the general solution. Next, we incorporate the effects
of the boundary conditions. The complementary (homogeneous) ODE corresponding
to the equation above is:
\begin{equation}
\frac{d^{2}}{dx_{0}^{2}}\overline{T}_{c}\left(x_{0}\right)+\varepsilon\frac{1-2x_{0}}{x_{0}\left(1-x_{0}\right)}\frac{d}{dx_{0}}\overline{T}_{c}\left(x_{0}\right)=0.
\end{equation}
The subscript $c$ on $\overline{T}_{c}$ indicates that this is the
complementary solution. By introducing a new variable 
\begin{equation}
\nu\left(x_{0}\right)=\frac{d\overline{T_{c}}}{dx_{0}},\label{eq:new-variable-intro}
\end{equation}
we can reduce the second-order ODE mentioned above to a first-order
ODE
\begin{equation}
\frac{1}{\nu\left(x_{0}\right)}d\nu\left(x_{0}\right)=-\varepsilon\frac{1-2x_{0}}{x_{0}\left(1-x_{0}\right)}dx_{0},\label{eq:NU-ODE}
\end{equation}
by integrating both sides of Eq.~\eqref{eq:NU-ODE}, we obtain the
following equality:
\begin{equation}
\ln\left(\nu\left(x_{0}\right)\right)=-\varepsilon\left(\ln\left(1-x_{0}\right)+\ln\left(x_{0}\right)\right)+\ln\left(c_{1}\right).
\end{equation}
By the exponentiation of the both sides of the above equality to eliminate
the logarithm, we express the variable $\nu\left(x_{0}\right)$ as:
\begin{equation}
\nu\left(x_{0}\right)=\frac{c_{1}}{x_{0}^{\varepsilon}\left(1-x_{0}\right)^{\varepsilon}}.\label{eq:Cnew-variable-defined}
\end{equation}
By integrating Eq.~\eqref{eq:new-variable-intro} with respect to
$x_{0}$, we find that the complementary solution can be expressed
in terms of $\nu\left(x_{0}\right)$:
\begin{equation}
\overline{T}_{c}\left(x_{0}\right)=c_{2}-c_{1}\int\nu\left(x_{0}\right)dx_{0}.\label{eq:Tc-integral-form}
\end{equation}
The integral on the right-hand side of $\overline{T}_{c}$ can be
expressed by using incomplete beta function 
\begin{equation}
\beta_{z}\left(a,b\right)=\int_{0}^{^{z}}r^{a-1}\left(1-r\right)^{b-1}dr,
\end{equation}
using this, the solution of the complementary ODE~\eqref{eq:ODE-gen-eps}
is:
\begin{equation}
\overline{T}_{c}\left(x_{0}\right)=c_{2}-c_{1}\beta_{1-x_{0}}\left(1-\varepsilon,1-\varepsilon\right).\label{eq:Complementary-solution}
\end{equation}
Here, $\beta_{1-x_{0}}\left(1-\varepsilon,1-\varepsilon\right)$ denotes
the incomplete beta function. While this function has singularities
at integer values of $\varepsilon$, as we will demonstrate later,
the integral in Eq.~\eqref{eq:Tc-integral-form} can still be evaluated
even when $\varepsilon$ takes integer values.

To find a particular solution $\overline{T}_{p}$ based on the complementary
solution $\overline{T}_{c}$, we apply the variation of parameters
method. According to this method, if the complementary solution of
the ODE is given by:
\begin{equation}
\overline{T}_{c}\left(x_{0}\right)=c_{1}s_{1}\left(x_{0}\right)+c_{2}s_{2}\left(x_{0}\right),\label{eq:Gen-complementary-solution}
\end{equation}
The particular solution can then be written as:
\begin{equation}
\overline{T}_{p}\left(x_{0}\right)=s_{2}\left(x_{0}\right)\int s_{1}\left(x_{0}\right)\frac{r\left(x_{0}\right)}{w\left(x_{0}\right)}dx_{0}-s_{1}\left(x_{0}\right)\int s_{2}\left(x_{0}\right)\frac{r\left(x_{0}\right)}{w\left(x_{0}\right)}dx_{0}.\label{eq:Gen-inhomogeneous-solution}
\end{equation}
Here, $w\left(x_{0}\right)$ denotes
\begin{equation}
w\left(x_{0}\right)=s_{1}\left(x_{0}\right)\frac{d}{dx_{0}}s_{2}\left(x_{0}\right)-s_{2}\left(x_{0}\right)\frac{d}{dx_{0}}s_{1}\left(x_{0}\right),
\end{equation}
where $r\left(x_{0}\right)$ represents the inhomogeneous part of
the ODE. From ODE~\eqref{eq:ODE-gen-eps}, it follows that in our
case:
\begin{equation}
r\left(x_{0}\right)=-\frac{1}{x_{0}\left(1-x_{0}\right)}.\label{eq:Inhom-part}
\end{equation}

By comparing Eq.~\eqref{eq:Complementary-solution} with Eq.~\eqref{eq:Gen-complementary-solution},
we observe that
\begin{equation}
s_{1}\left(x_{0}\right)=-\beta_{1-x_{0}}\left(1-\varepsilon,1-\varepsilon\right),\label{eq:S1(x)-sol-to-cont-1}
\end{equation}
and
\begin{equation}
s_{2}\left(x_{0}\right)=1.\label{eq:S2(x)-sol-to-cont-2}
\end{equation}
By substituting Eqs..~\eqref{eq:Inhom-part}, \eqref{eq:S1(x)-sol-to-cont-1},
and \eqref{eq:S2(x)-sol-to-cont-2} into Eq.~\eqref{eq:Gen-inhomogeneous-solution},
we obtain the particular solution: 
\begin{equation}
\overline{T}_{p}\left(x_{0}\right)=-\beta_{1-x_{0}}\left(1-\varepsilon,1-\varepsilon\right)\beta_{1-x_{0}}\left(\varepsilon,\varepsilon\right)-\int\left(\left(1-x_{0}\right)x_{0}\right)^{\varepsilon-1}\beta_{1-x_{0}}\left(1-\varepsilon,1-\varepsilon\right)dx_{0},\label{eq:particular-solution-in-the-form-of-beta-fuctions}
\end{equation}
in the integral form. The integral on the right-hand side of Eq.~\eqref{eq:particular-solution-in-the-form-of-beta-fuctions}
can be evaluated using the relation between the incomplete beta function
$B_{z}\left(a,b\right)$ and the hypergeometric function, defined
as:
\begin{equation}
\beta_{z}\left(a,b\right)=\bigg(\frac{z^{a}}{a}\bigg){}_{2}F_{1}\left(a,1-b;a+1;z\right),\label{eq:Gamma-hyper-relation}
\end{equation}
The function $_{2}F_{1}\left(a,1-b;a+1;z\right)$ is a Gauss hypergeometric
function, defined by the series:
\begin{equation}
_{2}F_{1}\left(a,1-b;a+1;z\right)=\sum_{k=0}^{\infty}\frac{\left(a\right)_{k}\left(1-b\right)_{k}}{\left(a+1\right)_{k}}\frac{z^{k}}{k!}.
\end{equation}
Here, $\left(a\right)_{k}$ is a Pochhammer symbol. By substituting
Eq.~\eqref{eq:Complementary-solution} into Eq..~\eqref{eq:particular-solution-in-the-form-of-beta-fuctions},
we obtain the particular solution:
\begin{equation}
\overline{T}_{p}\left(x_{0}\right)=\frac{1}{\Gamma\left(2\varepsilon\right)\Gamma\left(1-\varepsilon\right)}G_{3,3}^{2,3}\left(x_{0}\Bigg|\begin{array}{c}
1,1,2\left(1-\varepsilon\right)\\
1,1-\varepsilon,0
\end{array}\right),\label{eq:Particular-SOL}
\end{equation}
where, $\Gamma\left(\varepsilon\right)$ denotes the complete gamma
function, and $G_{p,q}^{m,n}\left(z\Bigg|\begin{array}{c}
a_{1},a_{2},..,a_{p}\\
b_{1},b_{2},..,b_{q}
\end{array}\right)$ is the Meijer~$G$-function.

The general solution of ODE~\eqref{eq:ODE-gen-eps} is given by
\begin{equation}
\overline{T}_{g}=\overline{T}_{c}+\overline{T}_{p},
\end{equation}
which is the sum of the complementary solution, Eq.~\eqref{eq:Gamma-hyper-relation},
and the particular solution, Eq.~\eqref{eq:Gamma-hyper-relation},
and can be expressed as:
\begin{equation}
\overline{T}_{g}\left(x_{0}\right)=c_{2}-c_{1}\beta_{1-x_{0}}\left(1-\varepsilon,1-\varepsilon\right)+\frac{1}{\Gamma\left(2\varepsilon\right)\Gamma\left(1-\varepsilon\right)}G_{3,3}^{2,3}\left(x_{0}\Bigg|\begin{array}{c}
1,1,2\left(1-\varepsilon\right)\\
1,1-\varepsilon,0
\end{array}\right).\label{eq:MFPT-gen}
\end{equation}
Using the boundary conditions, we determine the normalization constants
$c_{1}$ and$c_{2}$. We then derive an expression for the MFPT, $\overline{T}_{LH}$,
which explicitly incorporates the effects of the absorbing boundaries.
Recall that the noisy voter model is defined on the interval $x\in\left[0,1\right]$.
We begin by setting absorbing boundaries at points $L$ and $H$,
with $0\leq L<H\leq1$. As before, $L$ represents the lower boundary
and $H$ the upper boundary. The absorbing boundary conditions require
that the general solution of the ODE~\eqref{eq:ODE-gen-eps} satisfies
the following constraints:
\begin{equation}
\overline{T}_{g}\left(L\right)=0,\label{eq:Bound-L-MFPT}
\end{equation}
and
\begin{equation}
\overline{T}_{g}\left(H\right)=0.\label{eq:Bound-H-MFPT}
\end{equation}
Combining Eqs.~\eqref{eq:MFPT-gen}, \eqref{eq:Bound-L-MFPT}, and
\eqref{eq:Bound-H-MFPT}, we find that the MFPT with absorbing boundaries
at both ends is given by
\begin{equation}
\overline{T}_{LH}=\frac{\overline{T}_{p}\left(H\right)\beta_{1-L}-\overline{T}_{p}\left(L\right)\beta_{1-H}}{\beta_{1-H}-\beta_{1-L}}+\frac{\overline{T}_{p}\left(L\right)-\overline{T}_{p}\left(H\right)}{\beta_{1-H}-\beta_{1-L}}\beta_{1-x_{0}}+\overline{T}_{p}\left(x_{0}\right).\label{eq:Tvid_LH_AA}
\end{equation}
In the above $\beta_{z}=\beta_{z}\left(1-\varepsilon,1-\varepsilon\right)$
is the incomplete beta function (the frequently repeated arguments
have been omitted for brevity), and $\overline{T}_{p}\left(z\right)$
is a particular solution of Eq.~\eqref{eq:ODE-gen-eps}, which is
given by is given by Eq.~\eqref{eq:Particular-SOL}.

If we place a reflective boundary at $L$ at instead of using Eq~\eqref{eq:Bound-L-MFPT},
the general solution $\overline{T}_{g}\left(x_{0}\right)$ must satisfy:
\begin{equation}
\frac{d}{dx_{0}}\overline{T}_{g}\Big|_{x_{0}=L}=0.\label{eq:Reflective-cond-for-Tg}
\end{equation}
By substituting Eq.~\eqref{eq:MFPT-gen} into Eq.~\eqref{eq:Reflective-cond-for-Tg}
and differentiating with respect to $x_{0}$, we determine the first
normalization constant:
\begin{equation}
c_{1}\left(L\right)=\beta_{L}\left(\varepsilon,\varepsilon\right)-\frac{\Gamma\left(\varepsilon\right)\Gamma\left(\varepsilon\right)}{\Gamma\left(2\varepsilon\right)}.\label{eq:c1(L)-reflection-const}
\end{equation}
Next, inserting Eqs.~\eqref{eq:MFPT-gen} into Eq.~\eqref{eq:Bound-H-MFPT}
gives the second normalization constant:
\begin{equation}
c_{2}=c_{1}\left(L\right)\beta_{1-H}\left(1-\varepsilon,1-\varepsilon\right)-\overline{T}_{p}\left(H\right).\label{eq:c1(H)-abs-const}
\end{equation}
Inserting Eqs.~\eqref{eq:c1(L)-reflection-const} and ~\eqref{eq:c1(H)-abs-const}
into Eq.~\eqref{eq:MFPT-gen}, we obtain the MFPT for the scenario
with a reflective boundary at boundary at $L$ and absorbing boundary
at $H$: 
\begin{equation}
\overline{T}_{r}=\left(\frac{\Gamma\left(\varepsilon\right)\Gamma\left(\varepsilon\right)}{\Gamma\left(2\varepsilon\right)}-\beta_{L}\left(\varepsilon,\varepsilon\right)\right)\left(\beta_{1-x_{0}}-\beta_{1-H}\right)+\overline{T}_{p}\left(x_{0}\right)-\overline{T}_{p}\left(H\right).\label{eq:Tvid_LR_HA}
\end{equation}
For simplicity, we define $\overline{T}_{r}=\overline{T}_{L_{r}H_{a}}\left(x_{0}\right)$.
From this point on,$\overline{T}_{r}$ will always denote the MFPT
with a reflective boundary at the lower limit $L$ and an absorbing
boundary at the upper limit $H$.

The expressions for MFPT given in Eqs.~\eqref{eq:Tvid_LH_AA} and
~\eqref{eq:Tvid_LR_HA} are rather complex; however, we can derive
simpler approximations for large values of $\varepsilon$. When the
rate $\varepsilon$ is large, the influence of $x_{0}$ and $L$ in
Eq~\eqref{eq:Tvid_LR_HA} becomes negligible, allowing the MFPT to
be approximated by:
\begin{equation}
\overline{T}_{r}=-\frac{\Gamma\left(\varepsilon\right)\Gamma\left(\varepsilon\right)}{\Gamma\left(2\varepsilon\right)}\beta_{1-H}\left(1-\varepsilon,1-\varepsilon\right).\label{eq:T-r-large-eps}
\end{equation}
To simplify Eq~\eqref{eq:T-r-large-eps}, we use the following identity:
\begin{equation}
\frac{\Gamma\left(\varepsilon\right)\Gamma\left(\varepsilon\right)}{\Gamma\left(2\varepsilon\right)}=\beta\left(\varepsilon,\varepsilon\right),
\end{equation}
where is $\beta\left(a,b\right)$ is the Euler beta function, defined
by the integral
\begin{equation}
\beta\left(a,b\right)=\int_{0}^{1}r^{a-1}\left(1-r\right)^{b-1}dr.
\end{equation}
By applying the series expansion for large $\varepsilon$, the Euler
beta function can be approximated as
\begin{equation}
\beta\left(\varepsilon,\varepsilon\right)=\frac{2\sqrt{\pi}}{2^{2\varepsilon}\sqrt{\varepsilon}}+...,\label{eq:beta-large}
\end{equation}
and the incomplete beta function as
\begin{equation}
\beta_{1-H}\left(1-\varepsilon,1-\varepsilon\right)\simeq-\frac{1}{\varepsilon}\frac{H}{\left(1-H\right)^{-\varepsilon}H^{-\varepsilon}}.\label{eq:incomplete-beta-large}
\end{equation}
Substituting Eqs.~\eqref{eq:beta-large} and ~\eqref{eq:incomplete-beta-large}
into Eq.~\eqref{eq:T-r-large-eps}, we obtain: 
\begin{equation}
\overline{T}_{r}\simeq\frac{2\sqrt{\pi}}{2^{2\varepsilon}\varepsilon^{3/2}}\cdot\frac{H}{\left(1-H\right)^{\varepsilon}H^{\varepsilon}}=\frac{2\sqrt{\pi}H}{\varepsilon^{3/2}}\mathrm{e}^{-\varepsilon\ln\left[\frac{1}{4H\left(1-H\right)}\right]}.\label{eq:T-r-large-eps-approx}
\end{equation}
From Eq.~\eqref{eq:T-r-large-eps-approx}, it follows that the MFPT
with a reflective boundary at $L$ grows exponentially as the transition
rate $\varepsilon$ increases, i.e. $\overline{T}_{r}\sim\mathrm{e}^{\varepsilon}$.
A similar exponential growth can be demonstrated for the case with
two absorbing boundaries, where $\overline{T}_{LH}\sim\mathrm{e}^{\varepsilon}$.

Here, we derived an approximation for the MFPT with a reflective boundary
at $L$ and an absorbing boundary at $H$. This problem is equivalent
to the case where the reflection is at $H$ and absorption at $L$.
The results for this case could be obtained simply by swapping $L\rightarrow H$
and $H\rightarrow L$.

\subsection*{Mean first passage time for integer $\varepsilon$ values}

In the context of the noisy voter model, $\varepsilon$ represents
the independent transition rate and must be a non-negative real number.
Thus, the range of possible $\varepsilon$ values includes natural
numbers (non-negative integers). The general solutions for the noisy
voter model we have obtained in this work involve quite a few instances
of the incomplete beta function $\beta_{z}\left(a,b\right)$, which
is indeterminate when $a$ or $b$ is a negative integer. This is
an issue, as we often have that $a=b=1-\varepsilon$. Hence, for a
certain integer $\varepsilon$ values, the general solution we have
obtained would be indeterminate. However, these cases can be treated
independently by finding the solution directly from the ODE~\eqref{eq:ODE-gen-eps}.

Let us return to the general form of the complementary solution (see
Eq.~\eqref{eq:Tc-integral-form})
\begin{equation}
\overline{T}_{c}=c_{2}+c_{1}\int\frac{1}{x_{0}^{n}\left(1-x\right)_{0}^{n}}dx_{0}.\label{eq:Tc-gen-1}
\end{equation}
Here, we set $\varepsilon=n$. When $n$ is a natural number, integrating
the function
\begin{equation}
I\left(x\right)=\frac{1}{x^{n}\left(1-x\right)^{n}},
\end{equation}
requires breaking it down into partial fractions. To begin, we express
the decomposition as:
\begin{equation}
I\left(x\right)=\frac{1}{x^{n}\left(1-x\right)^{n}}=\sum_{i=1}^{n}\left[\frac{A_{i}\left(n\right)}{x^{i}}+\frac{B_{i}\left(n\right)}{\left(1-x\right)^{i}}\right],\label{eq:partial-fractions}
\end{equation}
where $A_{i}$ and $B_{i}$ are constants to be determined. These
coefficients are typically obtained by multiplying both sides of the
equation by $x^{n}\left(1-x\right)^{n}$ and matching coefficients
of like powers. Alternatively, one can evaluate suitable derivatives
at at $x=0$ and $x=1$. In practice, solving for the values of $A_{i}\left(n\right)$
and $B_{i}\left(n\right)$ becomes increasingly complex as $n$ grows,
because it involves solving a coupled system of linear equations that
scales with $n$. No simple closed-form expressions exist for $A_{i}\left(n\right)$
and $B_{i}\left(n\right)$, and the values vary across $n$; for example,
$A_{1}\left(1\right)\neq A_{1}\left(2\right)\neq\ldots\neq A_{1}\left(n\right)$.
Therefore, the coefficients must be computed case by case for each
specific integer value of $n$.

By substituting Eqs.~\eqref{eq:Tc-gen-1} and \eqref{eq:particular-solution-in-the-form-of-beta-fuctions}
(in place of Eqs.~\eqref{eq:Complementary-solution} and \eqref{eq:Particular-SOL})
into Eq.~\eqref{eq:Tvid_LH_AA}, we obtain an expression for the
MFPT that is valid for integer values of the transition rate.

For the scenario where both $L$ and $H$ are absorbing boundaries,
the corresponding MFPT is given by:
\begin{equation}
\overline{T}_{LH}=\frac{M_{n}\left(H\right)Q_{n}\left(L\right)-M_{n}\left(L\right)Q_{n}\left(H\right)}{Q_{n}\left(H\right)-Q_{n}\left(L\right)}+\frac{M_{n}\left(L\right)-M_{n}\left(H\right)}{Q_{n}\left(H\right)-Q_{n}\left(L\right)}Q_{n}\left(x_{0}\right)+M_{n}\left(x_{0}\right),\label{eq:T_HL_eps-N}
\end{equation}
where the auxiliary functions are defined as:
\begin{equation}
M_{n}\left(z\right)=\int^{z}\left(r\left(1-r\right)\right)^{n-1}\int^{r}\frac{1}{q^{n}\left(1-q\right)^{n}}dqdr-\int^{z}\frac{1}{r^{n}\left(1-r\right)^{n}}dr\int^{z}r^{n-1}\left(1-r\right)^{n-1}dr,\label{eq:f_N(z)}
\end{equation}
and
\begin{equation}
Q_{n}\left(z\right)=\int^{z}\frac{1}{r^{n}\left(1-r\right)^{n}}dr.\label{eq:g_N(z)}
\end{equation}

The MFPT for the case with a reflective boundary at $L$ and an absorbing
boundary at $H$ is given by:
\begin{equation}
\overline{T}_{r}=\left(\beta\left(n,n\right)-Q_{n}\left(L\right)\right)\left(Q_{n}\left(x_{0}\right)-Q_{n}\left(H\right)\right)+M_{n}\left(x_{0}\right)-M_{n}\left(H\right),\qquad\text{with }n\geq1,\label{eq:Tvid_LR_HA-1}
\end{equation}
the functions $M_{n}\left(z\right)$ and $Q_{n}\left(z\right)$ are
the same as in $\overline{T}_{LH}$, Eq.~\eqref{eq:T_HL_eps-N}.

\section{Special cases\label{sec:C}}

The general formulas for the MFPT, given in Eqs.~\eqref{eq:Tvid_LH_AA}
and~\eqref{eq:Tvid_LR_HA}, may at first seem difficult to evaluate
due to the presence of special functions. However, for certain values
of the transition rate $\varepsilon$, these expressions simplify
significantly. In particular, when $\varepsilon=\tfrac{n}{2}$ with
$n\in\mathbb{N}$, the MFPT can be written in terms of elementary
functions---specifically, trigonometric and logarithmic functions,
along with polynomials. For illustration, we present the special cases
$\varepsilon\in\left\{ 0,\tfrac{1}{2},1,\tfrac{3}{2}\right\} $. With
straightforward but careful algebra, similar formulas can be derived
for larger values of $n$.

\subsection*{Detailed derivation for the $\varepsilon=0$ case}

In the limiting case $\varepsilon=0$, the MFPT could be obtained
by substituting this value into Eqs.~\eqref{eq:T_HL_eps-N} and \eqref{eq:Tvid_LR_HA-1}.
While the general-case derivation of the MFPT is quite involved---as
detailed in~\ref{sec:A}---here
we provide a step-by-step derivation for the simplest case to aid
readers interested in understanding the calculation. Specifically,
when individualistic behavior is absent (i.e., $\varepsilon=0$),
the ODE~\eqref{eq:ODE-gen-eps} reduces to:
\begin{equation}
\frac{d^{^{2}}}{dx_{0}^{2}}\overline{T}\left(x_{0}\big|\varepsilon=0\right)=-\frac{1}{x_{0}\left(1-x_{0}\right)}.\label{eq:ODE-gen-eps-1}
\end{equation}
The ODE~\eqref{eq:ODE-gen-eps-1} can be solved by direct integration,
and the solution is 
\begin{equation}
\overline{T}\left(x_{0}\big|\varepsilon=0\right)=c_{1}+c_{2}x_{0}+\left(x_{0}-1\right)\ln\left(1-x_{0}\right)+x_{0}\ln x_{0}.\label{eq:MFPT-eps0-gen}
\end{equation}
The absorbing boundary conditions require that the general solution
of the ODE~\eqref{eq:MFPT-eps0-gen} satisfies the following constraints:
\begin{equation}
\overline{T}\left(L\big|\varepsilon=0\right)=0,\label{eq:Bound-L-MFPT-e-0}
\end{equation}
and
\begin{equation}
\overline{T}\left(H\big|\varepsilon=0\right)=0.\label{eq:Bound-H-MFPT-e-0}
\end{equation}

The normalization constants $c_{1}$ and $c_{2}$ can be determined
using the boundary conditions introduced above. Substituting Eq.~\eqref{eq:MFPT-eps0-gen}
into Eqs.~\eqref{eq:Bound-L-MFPT-e-0} and \eqref{eq:Bound-H-MFPT-e-0}
yield the following system of equations:
\begin{equation}
\left\{ \begin{array}{c}
c_{1}+c_{2}L+\left(L-1\right)\ln\left(1-L\right)+L\ln\left(L\right)=0,\\
c_{1}+c_{2}H+\left(H-1\right)\ln\left(1-H\right)+H\ln\left(H\right)=0.
\end{array}\right.\label{eq:LH-eq-e0}
\end{equation}
By solving the system of equations above, we obtain the normalization
constants:
\begin{equation}
c_{1}=\frac{Lf\left(H\right)-Hf\left(L\right)}{H-L},\quad\text{and}\quad c_{2}=\frac{f\left(L\right)-f\left(H\right)}{H-L}.\label{eq:C12-e-0}
\end{equation}
Here, we introduce a placeholder function
\begin{equation}
f\left(z\right)=\left(z-1\right)\ln\left(1-z\right)-z\ln\left(z\right),\label{eq:f(z)}
\end{equation}
which allows us to greatly simplify the expressions for $c_{1}$ and
$c_{2}$ expressions above. By substituting Eq.~\eqref{eq:C12-e-0}
into Eq.~\eqref{eq:MFPT-eps0-gen}, we obtain the MFPT for the case
with two absorbing boundaries:
\begin{equation}
\overline{T}_{LH}\left(x_{0}\big|\varepsilon=0\right)=\frac{Lf\left(L\right)-Hf\left(H\right)}{H-L}+\frac{f\left(L\right)-f\left(H\right)}{H-L}x_{0}+f\left(x_{0}\right).\label{eq:MFPT-eps-zero-HL}
\end{equation}

Let us set a reflective boundary at $L>0$, since from the Fokker--Planck
equation, Eq.~\eqref{eq:fokker-planck-nvm-x}, it is evident that
the diffusion coefficient becomes zero at the end points of the interval
$\left[0,1\right]$. In this special case, there is no drift, so even
if a reflective boundary is placed at the edge of the interval, there
is no noise or drift to alter the value of $x$. As a result, reflective
boundaries at the endpoints effectively behave as absorbing boundaries.

Therefore, to properly study the effect of a reflective boundary on
the MFPT, we must ensure that the reflective boundary is placed strictly
within the interval---that is, $L>0$. If we assume a reflective
boundary at point $L$, then instead of imposing $\overline{T}\left(L\right)=0$
, the condition from Eq.~\eqref{eq:C12-e-0} becomes:
\begin{equation}
\frac{d}{dx_{0}}\overline{T}\left(x_{0}\big|\varepsilon=0\right)\Bigg|_{x_{0}=L}=0.\label{eq:L-R-boundary-cond}
\end{equation}
By substituting Eq.~\eqref{eq:MFPT-eps0-gen} into Eq.~\eqref{eq:L-R-boundary-cond},
we obtain the normalization constant $c_{2}$:
\begin{equation}
c_{2}=\ln\left(L\right)-\ln\left(1-L\right),\label{eq:c2-L-R-eps-0}
\end{equation}
Then, by substituting Eq.~\eqref{eq:c2-L-R-eps-0} into the second
equation (concerning boundary $H$) from Eq.~\eqref{eq:LH-eq-e0},
we find the other constant:
\begin{equation}
c_{1}=H\left(\ln L-\ln\left(1-L\right)\right)-f\left(H\right).\label{eq:C1-RA-e-0}
\end{equation}

Here $f\left(z\right)$ is the same function as in the previous case
(see Eq.~\eqref{eq:f(z)}). By substituting Eqs.~\eqref{eq:c2-L-R-eps-0}
and \eqref{eq:C1-RA-e-0} into Eq.~\eqref{eq:MFPT-eps0-gen}, we
obtain MFPT for the case of reflective boundary at $L$
\begin{equation}
\overline{T}_{r}\left(x_{0}\big|\varepsilon=0\right)=\left(H-x_{0}\right)g\left(L\right)-f\left(H\right)+f\left(x_{0}\right).
\end{equation}
Here, we have introduce a new placeholder function, which is defined
as
\begin{equation}
g\left(z\right)=\ln\left(1-z\right)-\ln\left(z\right)=2\arctan\left(1-2z\right).
\end{equation}

\subsection*{Detailed derivation for the $\varepsilon=\tfrac{1}{2}$ case}

This case is particularly intriguing because it can be derived in
two distinct ways: either through our proposed approach or by applying
a nonlinear transformation to the Fokker--Planck equation, effectively
reducing it to a diffusion equation. We examine both methods in more
detail below.

The nonlinear transformation is especially useful for analyzing the
symmetry of the MFPT. However, for other values of $\varepsilon$,
it introduces a nonlinear drift term, leaving the transformed Fokker--Planck
equation just as complex as the original. Still, alternative transformations
may exist that simplify the voter model by mapping it onto well-known
stochastic processes---similar to techniques used in certain birth-death
models~\cite{Gontis2020PhysA}. This strategy has been successfully
applied to the scaled voter model to approximate the FPTD~\cite{Kazakevicius2023}.

In this case, we can use the relationship between the incomplete beta
function $\beta_{1-z}\left(a,b\right)$ inverse trigonometric functions:
\begin{equation}
\beta_{1-z}\left(1-\varepsilon,1-\varepsilon\right)\Big|_{\varepsilon=1/2}=\beta_{1-z}\left(1/2,1/2\right)=2\arcsin\left(\sqrt{1-z}\right).\label{eq:beta05}
\end{equation}
The Meijer~$G$-function can also be expressed in terms of inverse
trigonometric functions:
\begin{equation}
G_{3,3}^{2,3}\left(z\Bigg|\begin{array}{c}
1,1,1\\
1,1/2,0
\end{array}\right)=\frac{\sqrt{\pi}}{2}\left(\pi^{2}+\left(2\arccosh\left(\sqrt{z}\right)\right)^{2}\right).\label{eq:G05}
\end{equation}
By substituting Eq.~\eqref{eq:G05} into Eq.~\eqref{eq:Particular-SOL},
we obtain the corresponding particular solution.
\begin{equation}
\overline{T}_{p}\left(z\big|\varepsilon=\frac{1}{2}\right)=\frac{1}{2}\left(\pi^{2}+\left(2\arccosh\left(\sqrt{z}\right)\right)^{2}\right)\label{eq:Tp-eps05}
\end{equation}
Here we used the identity $\Gamma\left(1\right)\Gamma\left(\tfrac{1}{2}\right)=\sqrt{\pi}$.

By inserting Eqs.~\eqref{eq:beta05} and \eqref{eq:Tp-eps05} back
into Eq.~\eqref{eq:Tvid_LH_AA}, and applying properties of inverse
trigonometric and inverse hyperbolic functions, we obtain the MFPT
for absorbing boundaries:
\begin{equation}
\overline{T}_{LH}\left(x_{0}\big|\varepsilon=\frac{1}{2}\right)=\frac{1}{2}\left(\arcsin\left(2x_{0}-1\right)-\arcsin\left(2L_{x}-1\right)\right)\left(\arcsin\left(2H_{x}-1\right)-\arcsin\left(2x_{0}-1\right)\right).\label{eq:eq:Tvid_LH_AA-EPS-05}
\end{equation}

In the case of a reflective boundary at $L$, the formula for the
MFPT includes a special function---the Euler beta function. One important
identity to keep in mind is 
\begin{equation}
\beta\left(\frac{1}{2},\frac{1}{2}\right)=\pi.\label{eq:Eu-beta-at-0.5}
\end{equation}
To compute the MFPT, we substitute Eqs.~\eqref{eq:beta05}, \eqref{eq:Tp-eps05}
and \eqref{eq:Eu-beta-at-0.5} into Eq.~\eqref{eq:Tvid_LR_HA}. Then,
by applying properties of inverse hyperbolic functions, we arrive
at the final expression for the MFPT under this boundary condition:
\begin{align}
\overline{T}_{r}\left(x_{0}\big|\varepsilon=\frac{1}{2}\right) & =\frac{1}{2}\cdot\left(\arcsin\left(2H-1\right)-\arcsin\left(2x_{0}-1\right)\right)\nonumber \\
 & \quad\cdot\left(\arcsin\left(2H-1\right)-2\arcsin\left(2L-1\right)+\arcsin\left(2x_{0}-1\right)\right).\label{eq:Tvid_LH_RA-EPS-05-x-1}
\end{align}

An alternative approach to deriving the MFPT involves applying a variable
transformation to the Fokker--Planck equation or its corresponding
SDE. In the special case where $\varepsilon=\tfrac{1}{2}$, we can
use a nonlinear transformation of the form $y=\arcsin\left(2x-1\right)$.
By first applying this change of variables, the original Fokker--Planck
equation \eqref{eq:fokker-planck-nvm-x} transforms into a new equation
expressed in terms of the variable $y$:
\begin{equation}
\frac{\partial P\left(y,t\big|y_{0},0\right)}{\partial t}=\frac{\partial P\left(y,t\big|y_{0},0\right)}{\partial y^{2}},
\end{equation}
 and the corresponding time-backward equation has identical form
\begin{equation}
\frac{\partial P\left(y,t\big|y_{0},0\right)}{\partial t}=\frac{\partial P\left(y,t\big|y_{0},0\right)}{\partial y_{0}^{2}}.\label{eq:y-F-P-E}
\end{equation}

By comparing the equation above with the time-backward Fokker--Planck
equation, Eq.~\eqref{eq:Time-back-ward-F-P-E-AG}, and
using the previously established relation given in Fokker--Planck
equation, Eq.~\eqref{eq:MFPT-ODE-gen}, we arrive at the following
ODE for the MFPT:
\begin{equation}
\frac{d^{2}}{dy_{0}^{2}}\overline{T}\left(y_{0}\right)=-1.\label{eq:MFPT-eq}
\end{equation}
The ODE above can be solved by simple integration 
\begin{equation}
\overline{T}\left(y_{0}\right)=c_{1}+c_{2}y_{0}-\frac{y_{0}^{2}}{2}\label{eq:Tvid-Y-gen}
\end{equation}

The constants $c_{1}$ and $c_{2}$ can be determined by applying
the absorbing boundary conditions at the points $L_{y}$ and $H_{y}$.
These boundaries are introduced by enforcing that the MFPT satisfies
$\overline{T}\left(L_{y}\right)=0$ and $\overline{T}\left(H_{y}\right)=0$.
From these conditions, it follows that:
\begin{equation}
\left\{ \begin{array}{c}
\overline{T}\left(L_{y}\right)=c_{1}+c_{2}L_{y}-\frac{L_{y}^{2}}{2}=0,\\
\overline{T}\left(H_{y}\right)=c_{1}+c_{2}H_{y}-\frac{H_{y}^{2}}{2}=0.
\end{array}\right.
\end{equation}
From the system of equations above, we find that:
\begin{equation}
c_{1}=\frac{1}{2}H_{y}L_{y},\quad\text{and}\quad c_{2}=\frac{1}{2}\left(H_{y}+L_{y}\right).
\end{equation}

Substituting the values of $c_{1}$ and $c_{2}$ into Eq.~$\eqref{eq:Tvid-Y-gen}$,
we obtain the MFPT in the $y$-space corresponding to the boundaries
$L_{y}$ and $H_{y}$:
\begin{equation}
\overline{T}_{LH}\left(y_{0}\right)=\frac{1}{2}\left(y_{0}-L_{y}\right)\left(H_{y}-y_{0}\right).\label{eq:Tvid_LH_AA-EPS-05-Y}
\end{equation}
Recalling the relationship between the variables $y$ and $x$, given
by:
\begin{equation}
y=\arcsin\left(2x-1\right),\label{eq:y-to-x}
\end{equation}
and substituting it into Eq.~$\eqref{eq:Tvid_LH_AA-EPS-05-Y}$, we
obtain the MFPT in terms of $x$:
\begin{equation}
\overline{T}_{LH}\left(x_{0}\right)=\frac{1}{2}\left(\arcsin\left(2x_{0}-1\right)-\arcsin\left(2L-1\right)\right)\left(\arcsin\left(2H-1\right)-\arcsin\left(2x_{0}-1\right)\right)\label{eq:Tvid_LH_AA-EPS-05-x}
\end{equation}

We now turn to the symmetry properties of the MFPT in the presence
of absorbing boundaries. Our goal is to determine the location and
nature of the function's maximum, and to identify the conditions under
which the MFPT is symmetric about this point.

The MFPT in $y$-space, given in Eq.~$\eqref{eq:Tvid_LH_AA-EPS-05-Y}$,
is a quadratic function of the initial position $y_{0}$, and therefore
describes a parabola. Since parabolic functions are symmetric about
their vertex, we first identify the vertex location:
\begin{equation}
y_{\mathrm{0,max}}=\frac{H_{y}+L_{y}}{2}.\label{eq:y_Df(L,H)}
\end{equation}
To obtain the corresponding value of $x_{0}$, we invert the original
transformation:
\begin{equation}
x_{0}=\tfrac{1}{2}\left(1+\sin\left(y_{0}\right)\right).\label{eq:x_Df(y_0)}
\end{equation}
Substituting the vertex location from Eq.~\eqref{eq:y_Df(L,H)}
into Eq.~\eqref{eq:x_Df(y_0)}, we find that the maximum of
the MFPT occurs at:
\begin{equation}
x_{0,\mathrm{max}}=\tfrac{1}{2}\left(1+\sin\left(\tfrac{H_{y}+L_{y}}{2}\right)\right).
\end{equation}

We now examine whether the function is symmetric about this point.
That is, whether 
\begin{equation}
\overline{T}_{LH}\left(x_{0}\big|\varepsilon=\frac{1}{2}\right)=\overline{T}_{LH}\left(1-x_{0}\big|\varepsilon=\frac{1}{2}\right)
\end{equation}
holds. Since the transformation $x_{0}=\tfrac{1}{2}\left[1+\sin\left(y_{0}\right)\right]$
is both monotonic and smooth, any symmetry in $y_{0}$ directly translates
to symmetry in $x_{0}$. In particular, symmetry about the vertex
occurs when the quadratic is centered at $y_{0}=0$, which leads to
the condition:
\begin{equation}
H_{y}=-L_{y}\quad\Rightarrow\quad\arcsin\left(2H-1\right)=-\arcsin\left(2L-1\right).
\end{equation}
Applying the identity $\arcsin\left(-z\right)=-\arcsin\left(z\right)$,
we obtain the symmetry condition:
\begin{equation}
2H-1=-\left(2L-1\right),
\end{equation}
which simplifies to:
\begin{equation}
H=1-L.
\end{equation}
This means the interval $\left[L,H\right]$ is centered at $\tfrac{1}{2}$.
Any point$x_{0}$ within this interval has a mirror point $1-x_{0}$
that is equally distant from the midpoint. Because the boundaries
are placed symmetrically, the behavior of the stochastic process is
also symmetric, and so is the MFPT. This condition can be expressed
more explicitly by setting:
\begin{equation}
L=\frac{1}{2}-d,\quad H=\frac{1}{2}+d,\quad\text{with }0<d\leq\frac{1}{2}.
\end{equation}
Therefore, we conclude that the same symmetry condition holds for
the case $\varepsilon=\tfrac{1}{2}$ as for the case $\varepsilon=0$.

In the case of a reflecting boundary at $y_{0}=L_{y}$ and an absorbing
boundary at $y_{0}=H_{y}$, the constants $c_{1}$ and $c_{2}$ can
be determined by applying the boundary conditions:
\begin{equation}
\frac{d}{dy_{0}}\overline{T}\left(L_{y}\right)=0,\qquad\text{if }\overline{T}\left(H_{y}\right)=0.
\end{equation}
From the absorbing boundary condition, we have:
\begin{equation}
\overline{T}\left(H_{y}\right)=c_{1}+c_{2}H_{y}-\frac{H_{y}^{2}}{2}=0.
\end{equation}
This allows us to obtain the expression for the first constant:
\begin{equation}
c_{1}=\frac{1}{2}H_{y}\left(H_{y}-2c_{2}\right)
\end{equation}
Substituting $c_{1}$ back into Eq.~\eqref{eq:Tvid-Y-gen}, we obtain:
\begin{equation}
\overline{T}_{H}\left(y_{0}\right)=\frac{1}{2}H_{y}\left(H_{y}-2c_{2}\right)+c_{2}y_{0}-\frac{y_{0}^{2}}{2}.
\end{equation}
Applying the reflecting boundary condition:
\begin{equation}
\frac{d}{dy_{0}}\overline{T}_{H}\Big|_{y_{0}=L_{y}}=c_{2}-L_{y}=0,\quad\Rightarrow\quad c_{2}=L_{y}.
\end{equation}
Substituting $c_{2}$ back into Eq.~\eqref{eq:Tvid-Y-gen} , we arrive
at the final expression: 
\begin{equation}
\overline{T}_{r}\left(y_{0}\right)=\frac{1}{2}\left(H_{y}-y_{0}\right)\left(H_{y}-2L_{y}+y_{0}\right).\label{eq:Tvid_LH_RA-EPS-05-Y}
\end{equation}
Recalling the relationship between the variables $y$ and $x$, given
in Eq.~\eqref{eq:y-to-x} , we obtain the MFPT in $x$-space
\begin{align}
\overline{T}_{r}\left(x_{0}\right) & =\frac{1}{2}\cdot\left(\arcsin\left(2H-1\right)-\arcsin\left(2x_{0}-1\right)\right)\nonumber \\
 & \quad\cdot\left(\arcsin\left(2H-1\right)-2\arcsin\left(2L-1\right)+\arcsin\left(2x_{0}-1\right)\right).\label{eq:Tvid_LH_RA-EPS-05-x}
\end{align}
The obtained MFPT expressions, Eqs.~\eqref{eq:Tvid_LH_RA-EPS-05-Y}
and \eqref{eq:Tvid_LH_RA-EPS-05-x}, are identical to the expressions
obtained from the general solution, Eqs.~\eqref{eq:eq:Tvid_LH_AA-EPS-05}
and \eqref{eq:Tvid_LH_RA-EPS-05-x-1}.

\subsection*{Detailed derivation for the $\varepsilon=1$ case}

Let us derive the MFPT for the case where the transition rate parameter
is set to $\varepsilon=1$. For the case with two absorbing boundaries,
we demonstrate that the boundary values proposed in previous sections---namely
$L=\tfrac{1}{2}-d$ and $H=\tfrac{1}{2}+d$---result in an MFPT that
remains symmetric concerning the midpoint of the interval $\left[L,H\right]$.
We also consider the case when at $L$ we have reflective boundary
instead.

If we set $\varepsilon=1$, the integral in in Eq.~\eqref{eq:Tc-gen-1}
involves the kernel $\nu\left(x_{0}\right)$. Partial fraction decomposition
of the said kernel, according to to Eq.~\eqref{eq:partial-fractions}
, is given by
\begin{equation}
\frac{1}{x_{0}\left(1-x_{0}\right)}=\frac{1}{x_{0}}+\frac{1}{x_{0}\left(1-x_{0}\right)}.
\end{equation}
Thus the complementary solution is 
\begin{equation}
\overline{T}_{c}=c_{2}-c_{1}\left(\ln\left(x_{0}\right)-\ln\left(1-x_{0}\right)\right).\label{eq:Tc-e-1}
\end{equation}
Using the complementary solution $\overline{T}_{c}$ and applying
the method of variation of parameters, we obtain the particular solution
\begin{equation}
\overline{T}_{p}=\ln\left(1-x_{0}\right)\label{eq:eq:Tp-e-1}
\end{equation}
The general solution is given by $\overline{T}_{g}=\overline{T}_{c}+\overline{T}_{p},$
which is the sum of the complementary solution, Eq.~\eqref{eq:Tc-e-1},
and the particular solution, Eq.~\eqref{eq:eq:Tp-e-1}, and can be
expressed as:
\begin{equation}
\overline{T}_{g}=c_{2}-c_{1}\ln\left(\frac{x_{0}}{1-x_{0}}\right)+\ln\left(1-x_{0}\right).
\end{equation}
The presence of absorbing boundaries requires that $\overline{T_{g}}\left(L\right)=0$
and $\overline{T_{g}}\left(H\right)=0$. Therefore, the MFPT for two
absorbing boundaries is given by:
\begin{equation}
\overline{T}_{LH}\left(x_{0}\big|\varepsilon=1\right)=\frac{\ln\left(H\right)\ln\left(\frac{1-L}{1-x_{0}}\right)+\ln\left(L\right)\ln\left(\frac{1-x_{0}}{1-H}\right)+\ln\left(x_{0}\right)\ln\left(\frac{1-H}{1-L}\right)}{\ln\left(\frac{H}{1-H}\right)+\ln\left(\frac{L}{1-L}\right)}.\label{eq:MFPT-HL-eps-2}
\end{equation}

We can rewrite the MFPT in a more compact form:
\begin{equation}
\overline{T}_{LH}\left(x_{0}\big|\varepsilon=1\right)=\frac{1}{c_{LH}}\left(K\left(H,L,x_{0}\right)+K\left(L,x_{0},H\right)+K\left(x_{0},H,L\right)\right)\label{eq:T_HL-eps-2-K-form}
\end{equation}
Here, we define $K\left(a,b,c\right)=\ln\left(a\right)\ln\left(\tfrac{1-b}{1-c}\right)$,
and introduce the normalization factor as follows:
\begin{equation}
c_{LH}=\ln\left(\frac{H}{1-H}\right)+\ln\left(\frac{L}{1-L}\right)=2\left(\arctan\left(1-2H\right)-\arctan\left(1-2L\right)\right).
\end{equation}

The formula for the MFPT, given by Eq.~\eqref{eq:MFPT-HL-eps-2},
can be significantly simplified if we choose absorbing boundaries that
are symmetric concerning the the midpoint $\frac{1}{2}$ of the domain.
This symmetry means that the MFPT is invariant under the transformation
$x_{0}\to1-x_{0}$ when $\varepsilon=1$, i.e.,
\begin{equation}
\overline{T}_{LH}\left(x_{0}\big|\varepsilon=1\right)=\overline{T}_{LH}\left(1-x_{0}\big|\varepsilon=1\right).
\end{equation}
To make the notation more compact, we introduce the following logarithmic
variables:
\begin{equation}
\begin{aligned}A & =\ln\left(H\right), & B & =\ln\left(L\right), & C & =\ln\left(x_{0}\right),\\
D & =\ln\left(1-H\right), & E & =\ln\left(1-L\right), & F & =\ln\left(1-x_{0}\right).
\end{aligned}
\end{equation}
With these definitions, Eq.~\eqref{eq:MFPT-HL-eps-2} becomes
\begin{equation}
\overline{T}_{LH}\left(x_{0}\big|\varepsilon=1\right)=A\left(E-F\right)+B\left(F-D\right)+C\left(D-E\right)\label{eq:THL-A-F}
\end{equation}
The symmetry condition can be enforced by requiring
\begin{equation}
\overline{T}_{LH}\left(x_{0}\big|\varepsilon=1\right)-\overline{T}_{LH}\left(1-x_{0}\big|\varepsilon=1\right)=0,\label{eq:Symetry-condition}
\end{equation}
Substituting the Eq.~\eqref{eq:MFPT-HL-eps-2} into Eq.~\eqref{eq:Symetry-condition}
and simplifying leads to:
\begin{equation}
\left(A-B+D-E\right)\left(C-F\right)=0,
\end{equation}
This equation admits two possible solutions:
\begin{equation}
C-F=0\quad\Rightarrow\quad x=\frac{1}{2},
\end{equation}
and
\begin{equation}
A-B+D-E=0\quad\Rightarrow\quad\frac{H\left(1-H\right)}{L\left(1-L\right)}=1\quad\Rightarrow H=1-L.
\end{equation}
To express the symmetric-boundary case more explicitly, let us introduce
a small positive parameter $d\leq\tfrac{1}{2}$ that measures the
distance of the boundaries from the midpoint. Let us set that $L=\tfrac{1}{2}-d$.
Thus, form $H=1-L.$ follows the absorbing boundaries must satisfy:
\begin{equation}
L=\frac{1}{2}-d,\quad\Rightarrow\quad H=\frac{1}{2}+d.
\end{equation}

By inserting this parametrization into Eq.~\eqref{eq:MFPT-HL-eps-2},
we obtain a simplified closed-form expression for the MFPT under symmetric
absorbing boundaries.
\begin{align}
\overline{T}_{\frac{1}{2}-d,\frac{1}{2}+d}\left(x_{0}\big|\varepsilon=1\right) & =\frac{1}{c_{\frac{1}{2}-d,\frac{1}{2}+d}}\Bigg[\ln\left(\frac{1}{2}+d\right)\left\{ \ln\left(\frac{1}{2}+d\right)-\ln\left(x_{0}\left(1-x_{0}\right)\right)\right\} \nonumber \\
 & \quad+\ln\left(\frac{1}{2}-d\right)\left\{ \ln\left(x_{0}\left(1-x_{0}\right)\right)+\ln\left(\frac{1}{2}-d\right)\right\} \Bigg].
\end{align}

The MFPT with a reflective boundary at $L$, as given in Eq.~\eqref{eq:T_HL-eps-2-K-form},
can also be derived from Eq.~\eqref{eq:Tvid_LR_HA-1} by setting
$n=2$:
\begin{equation}
\overline{T}_{r}=-\left(1-L\right)\ln\left(1-H\right)-L\ln\left(H\right)+L\left(\ln\left(x_{0}\right)-\ln\left(1-x_{0}\right)\right)+\ln\left(1-x_{0}\right).
\end{equation}

\subsection*{Detailed derivation for the $\varepsilon=\tfrac{3}{2}$ case}

In this section, we derive the MFPT for a system with a transition
rate parameter set to $\varepsilon=\tfrac{3}{2}$. We consider two
configurations: one with absorbing boundaries and another with a reflective
boundary at the point $L$. For the case with two absorbing boundaries,
we demonstrate that the boundary values proposed in previous sections---namely
$L=\tfrac{1}{2}-d$ and $H=\tfrac{1}{2}+d$---result in an MFPT that
remains symmetric concerning the midpoint of the interval $\left[L,H\right]$.
In this case, we can use the relationship between the incomplete beta
function $\beta_{1-z}\left(a,b\right)$ inverse trigonometric functions:
\begin{equation}
\beta_{1-z}\left(-\frac{1}{2},-\frac{1}{2}\right)=\frac{2\left(1-2z\right)}{\sqrt{z\left(1-z\right)}}.\label{eq:beta15}
\end{equation}
The Meijer~$G$-function can also be expressed in terms of inverse
trigonometric functions:
\begin{equation}
G_{3,3}^{2,3}\left(z\Bigg|\begin{array}{c}
-1,1,1\\
-\frac{1}{2},1,0
\end{array}\right)=2\sqrt{\pi}\left(1+\frac{\left(1-2z\right)\arccos\left(\sqrt{z}\right)}{\sqrt{z\left(1-z\right)}}\right).\label{eq:Meijer-G-ant-e-1.5}
\end{equation}
 By substituting Eq.~\eqref{eq:Meijer-G-ant-e-1.5} into Eq.~\eqref{eq:Particular-SOL},
we obtain the corresponding particular solution:
\begin{equation}
\overline{T}_{p}\left(x_{0}\big|\varepsilon=\frac{3}{2}\right)=-\frac{1}{4\sqrt{\pi}}G_{3,3}^{2,3}\left(z\Bigg|\begin{array}{c}
-1,1,1\\
-\frac{1}{2},1,0
\end{array}\right)=\frac{1}{2}\left(\frac{\left(2x_{0}-1\right)\arccos\left(\sqrt{x_{0}}\right)}{\sqrt{x_{0}\left(1-x_{0}\right)}}-1\right),\label{eq:Tp-eps15}
\end{equation}
Here we used the gamma $\Gamma\left(z\right)$ function property:
\begin{equation}
\frac{1}{\Gamma\left(3\right)\Gamma\left(-\frac{1}{2}\right)}=-\frac{1}{4\sqrt{\pi}}
\end{equation}
By inserting Eqs.~\eqref{eq:beta15}, \eqref{eq:Tp-eps15} back into
Eq.~\eqref{eq:Tvid_LH_AA}, and applying properties of inverse trigonometric
and inverse hyperbolic functions, we obtain the MFPT for absorbing
boundaries:
\begin{align}
\overline{T}_{LH}\left(x_{0}\big|\varepsilon=\frac{3}{2}\right) & =\frac{1}{2\left(\frac{1-2H}{\sqrt{\left(1-H\right)H}}-\frac{1-2L}{\sqrt{\left(1-L\right)L}}\right)}\Bigg\{\frac{1-2H}{\sqrt{\left(1-H\right)H}}\left(1+\frac{\left(1-2L\right)\arccos\left(\sqrt{L}\right)}{\sqrt{\left(1-L\right)L}}\right)\nonumber \\
 & \quad-\frac{1-2L}{\sqrt{\left(1-L\right)L}}\left(1+\frac{\left(1-2H\right)\arccos\left(\sqrt{H}\right)}{\sqrt{\left(1-H\right)H}}\right)\nonumber \\
 & \quad+\frac{1-2x_{0}}{\sqrt{\left(1-x_{0}\right)x_{0}}}\left(\frac{\left(1-2H\right)\arccos\left(\sqrt{H}\right)}{\sqrt{\left(1-H\right)H}}-\frac{\left(1-2L\right)\arccos\left(\sqrt{L}\right)}{\sqrt{\left(1-L\right)L}}\right)\Bigg\}\nonumber \\
 & \quad-\frac{1}{2}\left(1+\frac{\left(1-2x_{0}\right)\arccos\left(\sqrt{x_{0}}\right)}{\sqrt{\left(1-x_{0}\right)x_{0}}}\right).\label{eq:MFPT-eps15-AA}
\end{align}

In the case of a reflective boundary at $L$, the formula for the
MFPT includes a special function---the Euler beta function. One important
identity to keep in mind is 
\begin{equation}
\beta\left(\frac{1}{2},\frac{1}{2}\right)=\frac{\pi}{8}.\label{eq:Eu-beta-at-0.5-1}
\end{equation}
To compute the MFPT, we substitute Eqs.~\eqref{eq:Eu-beta-at-0.5-1},
\eqref{eq:beta15} and \eqref{eq:Tp-eps15} into Eq.~\eqref{eq:Tvid_LR_HA}.
Then, by applying properties of inverse hyperbolic functions, we arrive
at the final expression for the MFPT under this boundary condition:
\begin{align}
\overline{T}_{r}\left(x_{0}\big|\varepsilon=\frac{3}{2}\right) & =\frac{1}{2}\left(1+\frac{\left(1-2H\right)\arccos\left(\sqrt{H}\right)}{\sqrt{\left(1-H\right)H}}\right)\nonumber \\
 & \quad+\frac{1}{8}\left(\frac{2\left(1-2x_{0}\right)}{\sqrt{\left(1-x_{0}\right)x_{0}}}-\frac{2\left(1-2H\right)}{\sqrt{\left(1-H\right)H}}\right)\left(\frac{2\left(1-2L\right)\left(1-L\right)L}{\sqrt{\left(1-L\right)L}}+2\arccos\left(\sqrt{L}\right)\right)\nonumber \\
 & \quad-\frac{1}{2}\left(1+\frac{\left(1-2x_{0}\right)\arccos\left(\sqrt{x_{0}}\right)}{\sqrt{\left(1-x_{0}\right)x_{0}}}\right).
\end{align}

However, the expressions for $\overline{T}_{LH}$ and $\overline{T}_{r}$
retain the same functional form as in Eq.~\eqref{eq:Tvid_LH_AA}.
They remain quite complex, and little simplification is possible unless
additional constraints are introduced. Let us assume that the boundaries
are placed symmetrically around the midpoint
\begin{equation}
L=\frac{1}{2}-d,\quad H=\frac{1}{2}+d,\qquad\text{with }0<d\leq\frac{1}{2}.\label{eq:delta-bound}
\end{equation}
By making this assumption, we arrive at a useful relationship:
\begin{equation}
\overline{T}_{p}\left(\frac{1}{2}+d\big|\varepsilon=\frac{3}{2}\right)-\overline{T}_{p}\left(\frac{1}{2}-d\big|\varepsilon=\frac{3}{2}\right)=\frac{\pi d}{\sqrt{1-4d}},\label{eq:Tp-delta}
\end{equation}
To obtain the relation above, we applied the identity:
\begin{equation}
\arccos\left(\sqrt{\frac{\pi}{2}+d}\right)+\arccos\left(\sqrt{\frac{\pi}{2}-d}\right)=\frac{\pi}{2}.
\end{equation}
By substituting Eqs.~\eqref{eq:delta-bound} and \eqref{eq:Tp-delta}
into Eq.~\eqref{eq:Tvid_LH_AA}, we derive the MFPT
\begin{align}
\overline{T}_{\frac{1}{2}-d,\frac{1}{2}+d}\left(x_{0}\big|\varepsilon=\frac{3}{2}\right) & =\frac{d}{\sqrt{1-4d^{2}}}\left(\arccos\left(\sqrt{\frac{1}{2}-d}\right)-\arccos\left(\sqrt{\frac{1}{2}+d}\right)\right)\nonumber \\
 & \quad+\frac{1-2x_{0}}{2\sqrt{\left(1-x_{0}\right)x_{0}}}\left(\frac{\pi}{4}-\arccos\left(\sqrt{x_{0}}\right)\right).
\end{align}
The first term ensures that $\overline{T}_{\frac{1}{2}-d,\frac{1}{2}+d}\left(x_{0}\big|\varepsilon=\tfrac{3}{2}\right)$
vanishes at the selected absorbing boundaries, while the second term
corresponds to the mean consensus time, $\overline{T}_{01}\left(x_{0}\big|\varepsilon=\tfrac{3}{2}\right)$.
Comparing this expression to the earlier Eq.~\eqref{eq:MFPT-eps15-AA},
we see that all terms responsible for the asymmetry have disappeared.
In principle, the same approach can be applied to any $\varepsilon=\tfrac{n}{2}$,
with $n\in\mathbb{N}$, since the MFPT expressions involve trigonometric,
logarithmic, and polynomial functions---all of which possess inherent
symmetries.

\end{singlespace}}

\end{document}